\documentclass[10pt,final,doublecolumn]{IEEEtran}
\usepackage{amsmath}
\usepackage{amssymb}
\usepackage{amsfonts}
\usepackage{latexsym}
\usepackage{graphicx}
\usepackage{bbding}
\usepackage{enumerate}
\usepackage{multicol}
\usepackage{color}
\usepackage{slashbox}
\usepackage{setspace}
\usepackage[ruled,linesnumbered]{algorithm2e}
\usepackage{bm}

\IEEEoverridecommandlockouts
\allowdisplaybreaks[4]

\begin{document}
%\begin{spacing}{1.65}

\title{Design of Non-Orthogonal Beamspace Multiple Access for Cellular Internet-of-Things}
\author{\authorblockN{
Rundong Jia, Xiaoming Chen, Caijun Zhong, Derrick Wing Kwan Ng, Hai Lin, and Zhaoyang Zhang
\thanks{Rundong Jia ({\tt 3130103623@zju.edu.cn}), Xiaoming Chen ({\tt chen\_xiaoming@zju.edu.cn}), Caijun Zhong ({\tt caijunzhong@zju.edu.cn}), and Zhaoyang Zhang ({\tt ning\_ming@zju.edu.cn}) are with the College of Information Science and Electronic Engineering, Zhejiang University, Hangzhou 310027, China. Derrick Wing Kwan Ng ({\tt w.k.ng@unsw.edu.au}) is with the School of Electrical Engineering and Telecommunications, the University of New South Wales, NSW 2052, Australia. Hai Lin ({\tt hai.lin@ieee.org}) is with the Department of Electrical and Information Systems, Graduate School of Engineering, Osaka Prefecture University, Sakai, Osaka 599-8531, Japan.}}}\maketitle

\begin{abstract}
In this paper, we study the problem of massive connections over limited radio spectrum for the cellular Internet-of-Things (IoT) in the fifth-generation (5G) wireless network. To address the challenging issues associated with channel state information (CSI) acquisition and beam design in the context of massive connections, we propose a new non-orthogonal beamspace multiple access framework. In particular, the user equipments (UEs) are non-orthogonal not only in the temporal-frequency domain, but also in the beam domain. We analyze the performance of the proposed non-orthogonal beamspace multiple access scheme, and derive an upper bound on the weighted sum rate in terms of channel conditions and system parameters. For further improving the performance, we propose three non-orthogonal transmit beam construction algorithms with different beamspace resolutions. Finally, extensive simulation results show that substantial performance gain can be obtained by the proposed non-orthogonal beamspace multiple access scheme over the baseline ones.
\end{abstract}


\section{Introduction}
The fast development of the Internet-of-Things (IoT) has triggered an exponential growth in the number of IoT devices to provide various emerging applications, e.g., metropolitan time-frequency perception, virtual navigation/virtual management, smart traffic, and environmental monitoring. Hence, as the most advanced wireless network, the fifth-generation (5G) wireless network is required to provide wireless access for a large number of IoT applications with certain quality of service (QoS) guarantee \cite{IoT}. In this context, the cellular IoT has been widely regarded as a basic function of the 5G wireless network \cite{CellularIoT1,CellularIoT2}.

Compared to the traditional cellular communications, the cellular IoT has several new characteristics, such as a massive number of access devices, a variety of application services, scarce wireless resources, and capability-constrained wireless devices. In other words, it is imperative to design a low-complexity massive access scheme for simple IoT devices with limited wireless resources. It is well known that conventional orthogonal multiple access (OMA) techniques, e.g., time division multiple access (TDMA) and frequency division multiple access (FDMA), exclusively allocate an orthogonal time-frequency resource block to a unique user equipment (UE). Although OMA techniques simplify the design of signal processing, they lead to a low spectral efficiency. As a result, it is difficult for OMA techniques to support massive access over limited radio spectrum. To this end, the non-orthogonal multiple access (NOMA) technique has been proposed as a viable solution to the cellular IoT \cite{MassiveNOMA1}-\cite{MassiveNOMA3}. In general, the NOMA technique employs superposition coding to realize massive spectrum sharing \cite{NOMA1,NOMA10}, and then utilizes successive interference cancellation (SIC) to mitigate the co-channel interference caused by non-orthogonal transmission \cite{NOMA2,NOMA3}. However, in the cellular IoT, the computational complexity of SIC that is proportional to the number of access devices might be unbearable for the capability-constrained IoT devices. To address this issue, the access devices are usually partitioned into several clusters, and SIC is only performed with respect to interference caused by the devices in the same cluster \cite{Clustering1,Clustering2}. However, user clustering leads to extra inter-cluster interference since SIC is performed only within a cluster. Therefore, the cluster-based NOMA should be deployed by combining with effective interference cancellation techniques. Considering multiple-antenna is a fundamental characteristic of 5G wireless networks, it is natural to apply spatial beamforming to mitigate the co-channel interference caused by NOMA \cite{5G1}. In particular, by exploiting the spatial degrees of freedom offered by the multiple-antenna systems, it is likely to jointly perform user clustering and interference cancellation in the spatial domain. As a simple example, the authors in \cite{MIMONOMA1} grouped the access devices into several clusters in the spatial domain, and the devices in a cluster share the same transmit beam. Then, spatial beamforming is performed to cancel the inter-cluster interference, and the SIC is adopted to eliminated the intra-cluster interference. In fact, the spatial degrees of freedom of multiple-antenna systems, i.e., the number of antennas equipped at the base station (BS), determines the capability of interference supression and the maximum number of supportable devices. In order to further improve the performance of massive access, the massive MIMO technique was applied to NOMA systems by deploying a large-scale antenna array at the BS \cite{MassiveMIMO1,MassiveMIMO2}.

To exploit the benefits brought by the spatial degrees of freedom for user clustering and interference cancellation in the cellular IoT, the multiple-antenna BS should have certain channel state information (CSI) about the access devices. In the ideal case, the BS can acquire full CSI by some means, and thus it is possible to perform optimal user clustering and cancel the inter-cluster interference effectively \cite{FullCSI1,FullCSI2}. However, it is not a trivial task to obtain CSI in multiple-antenna NOMA systems, since the BS is at the transmit side of the downlink channels. In practical systems, the CSI is usually obtained at the BS through quantized feedback in FDD mode \cite{CSIFeedback} and channel estimation in TDD mode \cite{CSIEstimation1}. However, in the scenario of massive access for the cellular IoT, the resources required for quantized feedback or channel estimation are prohibitive. To this end, a non-orthogonal channel estimation method for CSI acquisition was proposed in \cite{CSIEstimation2} to largely alleviate the burden in resource consumption. Yet, due to the co-channel interference, the non-orthogonal channel estimation method leads to a severe reduction of the CSI accuracy compared to the orthogonal one. Especially, the simple user clustering method based on channel spatial correlation is unfeasible and the commonly used interference cancellation method based on zero-forcing beamforming is inefficient in the case of low-precision CSI. As mentioned above, there is an emerging need to design a low-complexity massive access scheme for the cellular IoT. Recently, a beam division multiple access (BDMA) scheme was proposed for multiuser massive MIMO systems \cite{BDMA1,BDMA2}. Specifically, the users are simultaneously served by asymptotically orthogonal beams, which are constructed only based on the channel coupling matrix (namely channel correlation matrix). The orthogonal beams can effectively decrease the co-channel interference, and thus improve the overall performance. More importantly, channel coupling matrix varies over a relatively long period, and can be constructed with some slow time-varying parameters, e.g., angle of arrival (AoA) \cite{BDMA3,BDMA4}. Hence, BDMA is regarded as a simple and feasible multiple access scheme. Furthermore, if a beam of BDMA serves multiple users, namely non-orthogonal BDMA, the system is capable of supporting massive connections \cite{Beamspace1}. In other words, the clusters in the NOMA systems are separated in the beamspace, which also simplifies the design of user clustering \cite{Beamspace2}.

In the beamspace MIMO-NOMA systems, a beam is exploited to serve multiple devices in the same cluster simultaneously. Due to the random locations of the devices, a beam is unable to perfectly align all devices' channels. Therefore, although the beams are orthogonal, there still exists severe inter-cluster interference. In fact, orthogonal beamforming is generally suboptimal and redesigning the beamspace schemes is desired to improve the overall performance. In \cite{Beamspace3}, the authors proposed to adjust the beam direction by padding zeros in the beam domain, such that the beam can align the equivalent channel spanned by the devices. However, padding zeros is an integer programming problem, which is difficult to provide a general solution. To this end, we provide a new non-orthogonal beamspace multiple access framework for massive connections in the cellular IoT. In particular, the access devices are non-orthogonal not only in the temporary-frequency domain, but also in the beam domain. The non-orthogonal transmit beams can be directly constructed from the originally orthogonal base beams through linear combinations. The contributions of this paper are as follows:

\begin{enumerate}

\item This paper designs a comprehensive non-orthogonal beamspace multiple access scheme for the cellular IoT with massive connections, including CSI acquisition, user clustering, superposition coding, and SIC.

\item This paper analyzes the performance of the proposed non-orthogonal beamspace multiple access scheme and derives a closed-form expression for an upper bound on the weighted sum rate in terms of system parameters and channel conditions.

\item This paper optimizes the proposed non-orthogonal beamspace multiple access scheme, and presents three beam design algorithms with different degrees of freedom in the beamspace.

\end{enumerate}

The rest of this paper is organized as follows: Section II gives a brief introduction of the cellular IoT and presents a massive access framework in the beamspace. Section III first analyzes the weighted sum rate of the proposed massive access scheme, and then proposes three non-orthogonal transmit beams design algorithms. Next, Section IV provides extensive simulation results to validate the effectiveness of the proposed scheme. Finally, Section V concludes the paper.

\emph{Notations}: We use bold upper (lower) letters to denote matrices (column vectors), $(\cdot)^H$ to denote conjugate transpose, $\mathrm{E}[\cdot]$ to denote expectation, $\mathrm{var}(\cdot)$ to denote the variance, $\|\cdot\|$ to denote the $L_2$-norm of a vector, $\otimes$ to denote the Kronecker product, $\mathrm{vec}(\cdot)$ to denote the vectorization of a matrix, $|\cdot|$ to denote the absolute value, and $[x]^+=\max[x,0]$.

\section{System Model}
\begin{figure}[h] \centering
\includegraphics [width=0.48\textwidth] {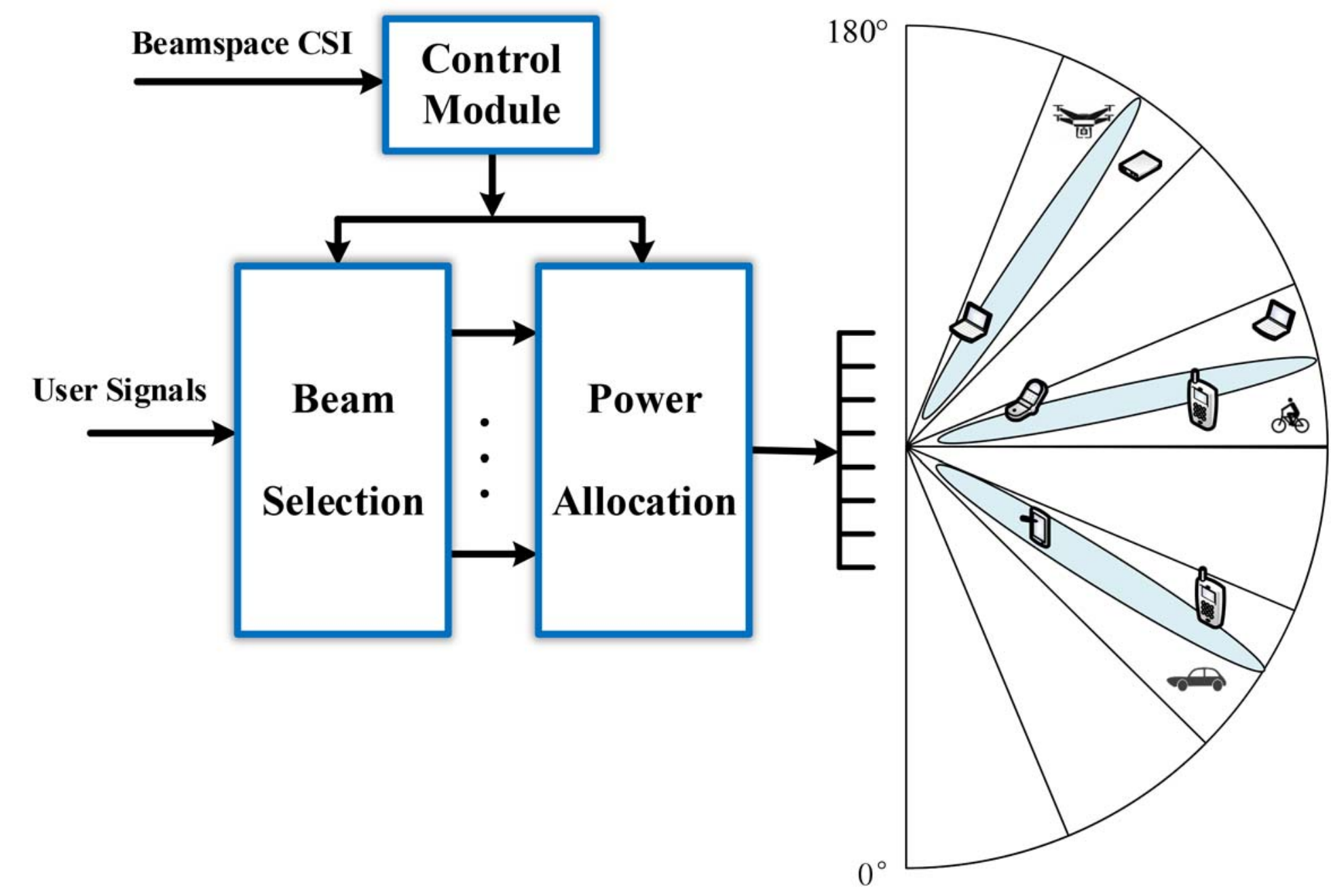}
\caption {A massive access model for the cellular IoT in the beamspace.}
\label{fig1}
\end{figure}

We consider a single-cell cellular IoT system as shown in Fig. \ref{fig1}\footnote{The proposed non-orthogonal beamspace multiple access scheme can be easily extended to the multiple-cell scenario.}, where a base station (BS) equipped with $N_t$ antennas broadcasts messages to $K$ IoT user equipments (UEs). Due to the size limitation, it is assumed that the IoT UEs have a single antenna each. Note that in the 5G cellular IoT, both the number of BS antennas $N_t$ and the number of UEs $K$ are very large \cite{IoT}. To support massive access over limited radio spectrum, we propose a non-orthogonal beamspace multiple access framework. In what follows, we introduce the key steps of non-orthogonal beamspace multiple access framework in detail.

\subsection{CSI Acquisition}
CSI availability at the BS is the precondition of designing the massive access scheme. However, in the context of massive connections, instantaneous CSI is hardly obtained due to the limited resources. To address this issue, we propose the use of long-term channel parameters, namely statistical CSI, to design the massive access scheme. We use $\mathbf{h}_k$ to denote the $N_t$-dimensional channel vector from the BS to the $k$th UE. It is assumed that the signal sent from the BS is scattered by a cluster and then arrives at the $k$th UE through $L$ propagation paths, where the $l$th path associated to the $k$th UE has a propagation distance $d_{k,l}$, a propagation attenuation $a_{k,l}$, and an angle of departure (AoD) $\theta_{k,l}$. Then, according to the electromagnetic wave propagation theory \cite{BDMA4}, \cite{Propagation}, the channel vector $\mathbf{h}_k$ can be expressed as
\begin{equation}
\mathbf{h}_k=\sum\limits_{l=1}^{L}a_{k,l}e^{-j2\pi d_{k,l}/\lambda_c}\mathbf{e}(\theta_{k,l}),\label{eqn1}
\end{equation}
where $\lambda_{c}$ is the carrier wavelength, and $\mathbf{e}(\theta_{k,l})$ is the transmit antenna array response vector. For a typical uniform linear array (ULA), $\mathbf{e}(\theta_{k,l})$ is given by
\begin{eqnarray}
\mathbf{e}(\theta_{k,l})&=&\frac{1}{\sqrt{N_t}}\left[1,\cdots,e^{-j2\pi(i-1)\varrho\sin(\theta_{k,l})},\right.\cdots,\nonumber\\
&&\left.e^{-j2\pi(N_t-1)\varrho\sin(\theta_{k,l})}\right]^T,\forall i\in[1,N_t],\label{eqn2}
\end{eqnarray}
where $\varrho$ is the normalized array spacing with respect to the carrier wavelength. %{\color{blue}And the transmit channel correlation matrix, which describe the channel statistical characteristics, can be described as:
%\begin{eqnarray}
%\mathbf{R}_k&=&\mathrm{E}\left[\mathbf{h}_k\mathbf{h}_k^H\right]\nonumber\\
%&=&\mathbf{U}_k\mathbf{\Lambda}_k\mathbf{U}_k^H.\label{eqn4}
%\end{eqnarray}
%}
%In the considered 5G cellular IoT, since the number of BS antennas is very large, the transmit antenna array response vectors are asymptotically orthogonal, cf. \cite{BDMA1} and \cite{BDMA2}.
%Thus, we can construct these vectors as a normalized matrix. With $N_t$ antennas, the BS can separate $N_t$ AoDs, namely $\phi_{i}, i=1,\cdots,N_t$. For the ULA, it is usual to adopt uniform sampling of spatial angles, i.e., $\sin\phi_{i}=\frac{2i}{N_t}-1$ with $\phi_{i}\in[-\frac{\pi}{2},\frac{\pi}{2}]$.}
Then, we construct these transmit antenna array response vectors as a matrix $\mathbf{U}_k=[\mathbf{e}(\phi_{k,1}),\mathbf{e}(\phi_{k,2}),\cdots,\mathbf{e}(\phi_{k,N_t})]\in\mathbb{C}^{N_t\times N_t}$. In this case, the channel vector $\mathbf{h}_k$ can be further transformed as
\begin{eqnarray}
\mathbf{h}_k&=&\sum\limits_{l=1}^{L}a_{k,l}e^{-j2\pi d_{k,l}/\lambda_c}\mathbf{e}(\theta_{k,l})\nonumber\\
&=&\sum\limits_{i=1}^{N_t}\mathbf{e}(\phi_{k,i})\sum\limits_{l\in \mathcal{L}_i}a_{k,l}e^{-j2\pi d_{k,l}/\lambda_c}\nonumber\\
&=&\mathbf{U}_k\mathbf{\Lambda}_k^\frac{1}{2}\bar{\mathbf{h}}_k,\label{eqn3}
\end{eqnarray}
%where $\mathbf{U}=[\mathbf{e}(\phi_{1}),\cdots,\mathbf{e}(\phi_{N_t})]$ denotes the transmit antenna array response matrix over $N_t$ orthogonal base beam directions,
where $\mathcal{L}_i$ is the set of all paths whose angles are equal to the angle $\phi_{k,i}$, and $\mathbf{\Lambda}_k=\mathrm{diag}\{\eta_{k,1},\cdots,\eta_{k,N_t}\}$ is a diagonal matrix with $\eta_{k,i}$ as the $i$th diagonal element being the gain in the angle $\phi_{k,i}$. In general, $\eta_{k,i}$ is determined by the sum of the propagation attenuation of multiple paths, whose angles are equal to the angle $\phi_{k,i}$. Moreover, $\bar{\mathbf{h}}_k$ is a $N_t$-dimensional complex Gaussian vector with zero mean and unit variance.

In the considered 5G cellular IoT, since the number of BS antennas is very large, we have the following lemma \cite{R1}, \cite{R2}:

\emph{Lemma 1}: If the number of BS antennas is sufficiently large, the beamspace matrix $\mathbf{U}_k$ is asymptotically identical for all the UEs, i,e.,
\begin{equation}
\mathbf{U}_k\to \mathbf{U}, \forall k,\quad \textrm{as}\,N_t\to \infty,\label{eqn4}
\end{equation}
where the $i$th column of $\mathbf{U}$, namely $\mathbf{u}_i$, can be expressed as $\mathbf{e}(\arcsin(\frac{2i}{N_t}-1))$, which means that the AoDs of all paths are sampled at some fixed angles. If the number of BS antennas is sufficiently large, the error caused by angle sampling can be ignored. Specifically, assuming the BS has $N_t$ antennas, the maximum angle sampling error is $\frac{\pi}{2N_t}$ and the average relative error is $\frac{1}{4N_t}$. For example, when the number of BS antennas is $N_t=64$, the average relative error is 0.39\%, which is negligible.

Furthermore, the channel vector $\mathbf{h}_k$ and the transmit channel correlation matrix $\mathbf{R}_k$ can be expressed as
\begin{equation}
\mathbf{h}_k=\mathbf{U}\mathbf{\Lambda}_k^\frac{1}{2}\bar{\mathbf{h}}_k,\label{eqn5}
\end{equation}
and
\begin{eqnarray}
\mathbf{R}_k&=&\mathrm{E}\left[\mathbf{h}_k\mathbf{h}_k^H\right]\nonumber\\
&=&\mathbf{U}_k\mathbf{\Lambda}_k\mathbf{U}_k^H,\label{eqn}
\end{eqnarray}
respectively. In other words, $\mathbf{h}_k$ can be represented as a linear combination of $N_t$ base beam directions. All the combinations of these orthogonal base beams form a beamspace.

Given the channel correlation matrix, it is possible to obtain the gain matrix $\mathbf{\Lambda}_k$ over the $N_t$ base beam directions. In fact, the channel correlation matrix remains unchanged during a relative long time and is easily obtained by statistical averaging over a large number of channel realizations. Thus, we can get the gains of the beam directions based on statistical CSI. In summary, the BS is capable of obtaining the beamspace CSI, i.e., the base beams $\mathbf{u}_i$ and the corresponding gains $\eta_{k,i}$ of all the UEs, $\forall i\in[1,N_t], k\in[1,K]$.

\subsection{User Clustering}
User clustering can effectively reduce the computational complexity at the UEs as the SIC is only performed within a cluster, which is a critical issue for the IoT devices. It is intuitive that user clustering should be performed based on available CSI at the BS. In the proposed framework, the BS only utilizes the statistical CSI or beamspace CSI to simplify the system complexity. Accordingly, we propose to perform user clustering in the beamspace. Generally speaking, the beamspace $[-\frac{\pi}{2},\frac{\pi}{2}]$ is divided into $N_t$ subspaces and a subspace corresponds to a cluster\footnote{Note that the subspace is determined by the angular resolution of the large-scale antenna array. Thus, the maximum number of clusters is $N_t$, but we also can use multiple subspaces to form a cluster. As a result, the number of clusters can be decreased.}. Therefore, an UE whose AoD belongs to $[-\frac{\pi}{2}+(i-1)\frac{\pi}{N_t},-\frac{\pi}{2}+i\frac{\pi}{N_t}]$ is grouped into the $i$th cluster. For the AoD information of the UEs, it can be obtained by using existing algorithms, e.g., the MUSIC algorithm \cite{MUSIC}. Hence, we can easily partition the UEs into multiple clusters according to the AoD information. Due to the random distribution of the IoT UEs, we assume that there are $M$ clusters\footnote{The system may have $N_t$ clusters at most, but only $M$ clusters are non-empty. Thus, we have $1\leq M\leq N_t.$} and $N_m$ UEs in the $m$th cluster. For ease of notation, we use UE$_{m,n}$ and $\mathbf{h}_{m,n}$ to denote the $n$th UE in the $m$th cluster and the corresponding channel vector, respectively.

\subsection{Superposition Coding}
\begin{figure}[h] \centering
\includegraphics [width=0.48\textwidth] {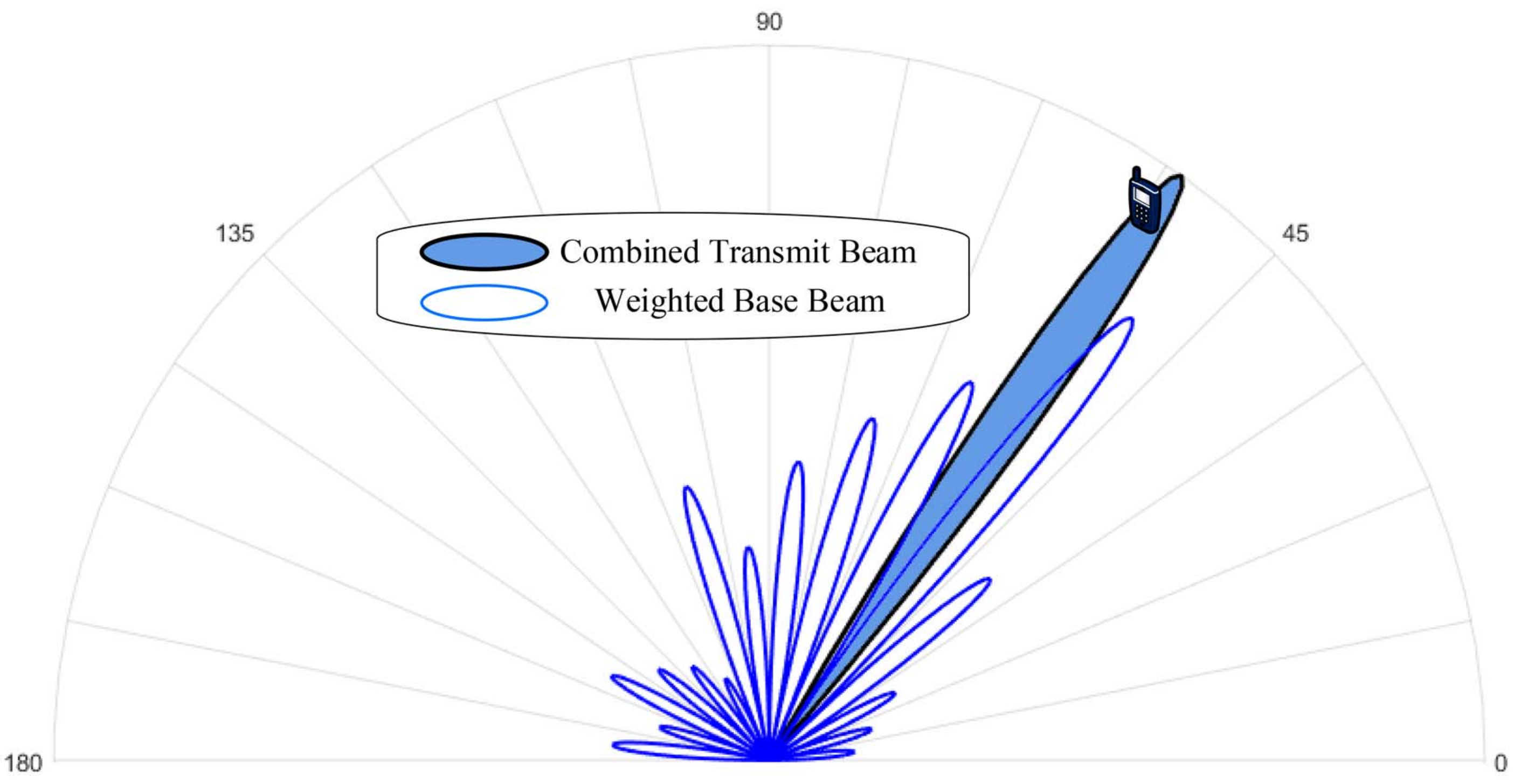}
\caption {The construction of the non-orthogonal transmit beam.}
\label{fig2}
\end{figure}

Superposition coding at the BS is a key step for achieving efficient massive access over limited radio spectrum. In general, superposition coding can be regarded as a weighted sum of the signals to be transmitted, which is mathematically given by
\begin{equation}
\mathbf{x}=\sum\limits_{m=1}^{M}\sum\limits_{n=1}^{N_m}\mathbf{w}_{m,n}x_{m,n},\label{eqn6}
\end{equation}
where $x_{m,n}$ and $\mathbf{w}_{m,n}$ is the Gaussian distributed transmit signal with unit norm and the transmit beam for the UE$_{m,n}$, respectively. In previous related works \cite{Beamspace1} and \cite{Beamspace2}, the transmit beam is directly designed based on the transmit antenna array vector associated to its assigned cluster. For instance, since the UE$_{m,n}$ falls in the $m$th cluster, $\mathbf{w}_{m,n}$ can be simply constructed as $\mathbf{u}_i$, namely matching filter (MF) beamforming. An advantage of such a method is that the beams across the clusters are orthogonal of each other if the number of BS antennas is sufficiently large. However, due to the limited angular resolution, a beam cannot perfectly align multiple UEs in a cluster. Hence, the co-channel interference does exist even the MF beamforming is performed. To solve this problem, we propose a non-orthogonal beamspace multiple access framework. As shown in Fig. \ref{fig2}, a transmit beam is a linear combination of multiple transmit antenna array vectors, namely base beams. In particular, by optimizing the weighted coefficients, the non-orthogonal beamspace multiple access scheme can significantly increase the angular resolution, decrease the inter-beam interference, and thus improves the overall performance of the cellular IoT. In this case, the transmit beam for the UE$_{m,n}$ can be expressed as
\begin{eqnarray}
\mathbf{w}_{m,n}&=&\sum\limits_{c\in\mathcal{B}_{m,n}}\sqrt{p_{m,n,c}}\mathbf{u}_c\nonumber\\
&=&\mathbf{U}\mathbf{P}_{m,n}^\frac{1}{2}\mathbf{s}_{m,n},\label{eqn7}
\end{eqnarray}
where $\mathcal{B}_{m,n}$ is the index set of selected base beams for the UE$_{m,n}$, $p_{m,n,c}$ is the transmit power on the $c$th base beam, and $\mathbf{P}_{m,n}=\mathrm{diag}\{p_{m,n,1},\cdots,p_{m,n,N_t}\}$. Moreover, $\mathbf{s}_{m,n}=[s_{m,n,1},\cdots,s_{m,n,N_t}]^T$ is the beam selection vector. If $i\in\mathcal{B}_{m,n}$, then $s_{m,n,i}=1$, otherwise $s_{m,n,i}=0$.  As such, superposition coding is equivalent to the design of non-orthogonal transmit beams, namely a combination of beam selection and power allocation, which will be discussed in detail later.

\subsection{Successive Interference Cancellation}
With the superposition coded signal $\mathbf{x}$, the BS broadcasts it over the downlink channels. Without loss of generality, we consider the received signal at the UE$_{m,n}$, which is given by
\begin{eqnarray}
y_{m,n}&=&\mathbf{h}_{m,n}^H\mathbf{x}+n_{m,n}\nonumber\\
&=&{\underbrace{\mathbf{h}^H_{m,n}\mathbf{w}_{m,n}x_{m,n}}_{\textrm{Desired signal}}}+{\underbrace{\sum\limits_{i=1,i\neq n}^{N_m}\mathbf{h}^H_{m,n}\mathbf{w}_{m,i}x_{m,i}}_{\textrm{Intra-cluster interference}}}\nonumber\\
&&+{\underbrace{\sum\limits_{j=1,j\neq m}^{M}\sum\limits_{i=1}^{N_j}\mathbf{h}^H_{m,n}\mathbf{w}_{j,i}x_{j,i}}_{\textrm{Inter-cluster
interference}}}+{\underbrace{n_{m,n}}_{\mathrm{AWGN}}},\label{eqn8}
\end{eqnarray}
where $n_{m,n}$ is additive white Gaussian noise (AWGN) with unit variance. Note that the first term at the right side of Eq. (\ref{eqn8}) is the desired signal, the second one is the intra-cluster interference, and the third one is the inter-cluster interference. In order to further improve the quality of the received signal, the UE$_{m,n}$ carries out SIC within the associated cluster. It is assumed that the desired channel gains in the $m$th cluster has a descending order as follows:
\begin{equation}
g_{m,1}\geq\cdots g_{m,n}\geq g_{m,N_m},\label{eqn9}
\end{equation}
where $g_{m,n}$ is defined as
\begin{eqnarray}
g_{m,n}&=&\|\mathbf{h}_{m,n}\|^2\nonumber\\
&=&\bar{\mathbf{h}}_{m,n}^H\mathbf{\Lambda}_{m,n}^\frac{1}{2}\mathbf{U}^H\mathbf{U}\mathbf{\Lambda}_{m,n}^\frac{1}{2}\bar{\mathbf{h}}_{m,n}\nonumber\\
&=&\bar{\mathbf{h}}_{m,n}^H\mathbf{\Lambda}_{m,n}\bar{\mathbf{h}}_{m,n},\label{eqn10}
\end{eqnarray}
with $\bar{\mathbf{h}}_{m,n}$ being the small-scale channel fading vector, and $\mathbf{\Lambda}_{m,n}=\mathrm{diag}\{\eta_{m,n,1},\cdots,\eta_{m,n,N_t}\}$ being the gain matrix of the orthogonal base beams for the UE$_{m,n}$. In practice, each UE knows its channel gain through channel estimation for coherent detection \cite{CoherentDetection}. Thus, the UEs can feed back the channel gains to the BS via the uplink and the BS determines the order of channel gains in each cluster which is informed to the UEs through the downlink. Based on the order of the channel gains, the UE$_{m,n}$ decodes the interfering signals and subtracts the corresponding interference from the $N_m$th to $(n+1)$th UE in sequence, and finally demodulates its desired signal $x_{m,n}$ \cite{R3}. Hence, the received signal after SIC at the UE$_{m,n}$ can be expressed as
\begin{eqnarray}
y_{m,n}'&=&\mathbf{h}^H_{m,n}\mathbf{w}_{m,n}x_{m,n}+\sum\limits_{i=1}^{n-1}\mathbf{h}^H_{m,n}\mathbf{w}_{m,i}x_{m,i}\nonumber\\
&&+\sum\limits_{j=1,j\neq m}^{M}\sum\limits_{i=1}^{N_j}\mathbf{h}^H_{m,n}\mathbf{w}_{j,i}x_{j,i}+n_{m,n}.\label{eqn11}
\end{eqnarray}
Thus, the corresponding signal-to-interference-plus-noise ratio (SINR) and the achievable rate of the UE$_{m,n}$ can be computed as (\ref{eqn12}) and (\ref{eqn13}) at the top of the next page.
\begin{figure*}
\begin{eqnarray}
\gamma_{m,n}&=&\frac{|\mathbf{h}^H_{m,n}\mathbf{w}_{m,n}|^2}{\sum\limits_{i=1}^{n-1}|\mathbf{h}^H_{m,n}\mathbf{w}_{m,i}|^2+\sum\limits_{j=1,j\neq m}^{M}\sum\limits_{i=1}^{N_j}|\mathbf{h}^H_{m,n}\mathbf{w}_{j,i}|^2+1}\nonumber\\
&=&\frac{|\bar{\mathbf{h}}_{m,n}^H\mathbf{\Lambda}_{m,n}^\frac{1}{2}\mathbf{P}_{m,n}^\frac{1}{2}\mathbf{s}_{m,n}|^2}{\sum\limits_{i=1}^{n-1}|\bar{\mathbf{h}}_{m,n}^H\mathbf{\Lambda}_{m,n}^\frac{1}{2}\mathbf{P}_{m,i}^\frac{1}{2}\mathbf{s}_{m,i}|^2+\sum\limits_{j=1,j\neq m}^{M}\sum\limits_{i=1}^{N_j}|\bar{\mathbf{h}}_{m,n}^H\mathbf{\Lambda}_{m,n}^\frac{1}{2}\mathbf{P}_{j,i}^\frac{1}{2}\mathbf{s}_{j,i}|^2+1},\label{eqn12}
\end{eqnarray}

\begin{eqnarray}
r_{m,n}&=&\log_2(1+\gamma_{m,n})\nonumber\\
&=&\log_2\left(\frac{\sum\limits_{j=1,j\neq m}^{M}\sum\limits_{i=1}^{N_j}|\bar{\mathbf{h}}_{m,n}^H\mathbf{\Lambda}_{m,n}^\frac{1}{2}\mathbf{P}_{j,i}^\frac{1}{2}\mathbf{s}_{j,i}|^2+\sum\limits_{i=1}^{n}|\bar{\mathbf{h}}_{m,n}^H\mathbf{\Lambda}_{m,n}^\frac{1}{2}\mathbf{P}_{m,i}^\frac{1}{2}\mathbf{s}_{m,i}|^2+1}
{\sum\limits_{j=1,j\neq m}^{M}\sum\limits_{i=1}^{N_j}|\bar{\mathbf{h}}_{m,n}^H\mathbf{\Lambda}_{m,n}^\frac{1}{2}\mathbf{P}_{j,i}^\frac{1}{2}\mathbf{s}_{j,i}|^2+\sum\limits_{i=1}^{n-1}|\bar{\mathbf{h}}_{m,n}^H\mathbf{\Lambda}_{m,n}^\frac{1}{2}\mathbf{P}_{m,i}^\frac{1}{2}\mathbf{s}_{m,i}|^2+1}\right).\label{eqn13}
\end{eqnarray}
\end{figure*}
From (\ref{eqn13}), it is known that the achievable rate is mainly determined by the beam selection and the corresponding power allocation, namely the design of non-orthogonal transmit beams. To improve the system performance, we design the non-orthogonal beamforming algorithms according to the characteristics of the beamspace massive access system.

\section{Performance Analysis and Optimization of Non-Orthogonal Beamspace Multiple Access}
In this section, we aim to design the non-orthogonal beamspace multiple access scheme from the perspective of maximizing the weighted sum of the ergodic rates of the cellular IoT. To facilitate the design, we first analyze the weighted sum of the ergodic rates.

\subsection{Performance Analysis}
In general, the weighted sum of the ergodic rate can be expressed as
\begin{equation}
R_{\mathrm{sum}}=\sum\limits_{m=1}^{M}\sum\limits_{n=1}^{N_m}\alpha_{m,n}\mathrm{E}[r_{m,n}],\label{eqn14}
\end{equation}
where $\alpha_{m,n}>0$ denotes the priority of the UE$_{m,n}$. As seen in (\ref{eqn13}), $r_{m,n}$ is a complicated function of the random variable $\|\bar{\mathbf{h}}_{m,n}\|^2$, thus it is difficult to obtain a closed-form expression for the weighted sum rate $R_{\mathrm{sum}}$. As an alternative, we derive an upper bound on the weighted sum of the ergodic rates, which is summarized in the following theorem:

\emph{Theorem 1}: Based on the proposed non-orthogonal beamspace multiple access framework, the weighted sum of the ergodic rates of all UEs is upper bounded by (\ref{eqn15}) at the top of the next page.
\begin{figure*}
\begin{eqnarray}
R_{\mathrm{ub}}&=&\sum\limits_{m=1}^{M}\sum\limits_{n=1}^{N_m}\sum\limits_{c=1}^{N_t}\alpha_{m,n}\left(\log_2\left(\left(\sum\limits_{j=1,j\neq m}^{M}\sum\limits_{i=1}^{N_j}s_{j,i,c}p_{j,i,c}
+\sum\limits_{i=1}^{n}s_{m,i,c}p_{m,i,c}\right)\eta_{m,n,c}+1\right)\right.\nonumber\\
&&\left.-\log_2\left(\left(\sum\limits_{j=1,j\neq m}^{M}\sum\limits_{i=1}^{N_j}s_{j,i,c}p_{j,i,c}+\sum\limits_{i=1}^{n-1}s_{m,i,c}p_{m,i,c}\right)\eta_{m,n,c}+1\right)\right)\label{eqn15}\\
&=&\sum\limits_{m=1}^{M}\sum\limits_{n=1}^{N_m}\sum\limits_{c=1}^{N_t}\alpha_{m,n}\log_2\left(1+\frac{s_{m,n,c}p_{m,n,c}\eta_{m,n,c}}{\sum\limits_{j=1,j\neq m}^{M}\sum\limits_{i=1}^{N_j}s_{j,i,c}p_{j,i,c}\eta_{m,n,c}+\sum\limits_{i=1}^{n-1}s_{m,i,c}p_{m,i,c}\eta_{m,n,c}+1}\right).\label{eqn16}
\end{eqnarray}
\end{figure*}
%\begin{eqnarray}
%R_{\mathrm{ub}}&=&\sum\limits_{m=1}^{M}\sum\limits_{n=1}^{N_m}\sum\limits_{c=1}^{N_t}\alpha_{m,n}\left(\log_2\left(\left(\sum\limits_{j=1,j\neq m}^{M}\sum\limits_{i=1}^{N_j}s_{j,i,c}p_{j,i,c}\right.\right.\right.\nonumber\\
%&&\,\,\,\left.\left.+\sum\limits_{i=1}^{n}s_{m,i,c}p_{m,i,c}\right)\eta_{m,n,c}+1\right)\nonumber\\
%&&-\log_2\left(\left(\sum\limits_{j=1,j\neq m}^{M}\sum\limits_{i=1}^{N_j}s_{j,i,c}p_{j,i,c}\right.\right.\nonumber\\
%&&\,\,\,\left.\left.\left.+\sum\limits_{i=1}^{n-1}s_{m,i,c}p_{m,i,c}\right)\eta_{m,n,c}+1\right)\right).\label{eqn15}
%\end{eqnarray}
\begin{proof}
Please refer to Appendix A.
\end{proof}
Note that the upper bound $R_{\mathrm{ub}}$ can be rewritten as (\ref{eqn16}). It is seen that the upper bound $R_{\mathrm{ub}}$ is equivalent to a sum rate on $N_t$ orthogonal resource blocks. As such, there is no interference among these resource blocks. Furthermore, according to the upper bound, we can obtain the following proposition:

\emph{Proposition 1}: In the scenario of massive connections, the weighted sum of the ergodic rates will be saturated.

\begin{proof}
Let $p_{j,i,c}=\nu_{j,i,c}P_{\mathrm{tot}}$, where $P_{\mathrm{tot}}$ is the total transmit power of the BS, and $\nu_{j,i,c}$ is the power allocation factor for the UE$_{j,i}$ over the $c$th base beam. In the scenario of massive connections, if the transmit power is high enough, the cellular IoT is usually interference-limited. In other words, the noise is negligible compared to the interference. Then, the upper bound is reduced to (\ref{eqn17}) at the top of the next page.
\begin{figure*}
\begin{eqnarray}
R_{\mathrm{ub}}&\approx&\sum\limits_{m=1}^{M}\sum\limits_{n=1}^{N_m}\sum\limits_{c=1}^{N_t}\alpha_{m,n}\log_2\left(1+\frac{s_{m,n,c}\nu_{m,n,c}P_{\mathrm{tot}}\eta_{m,n,c}}{\sum\limits_{j=1,j\neq m}^{M}\sum\limits_{i=1}^{N_j}s_{j,i,c}\nu_{j,i,c}P_{tot}\eta_{m,n,c}+\sum\limits_{i=1}^{n-1}s_{m,i,c}\nu_{m,i,c}P_{\mathrm{tot}}\eta_{m,n,c}}\right)\nonumber\\
&=&\sum\limits_{m=1}^{M}\sum\limits_{n=1}^{N_m}\sum\limits_{c=1}^{N_t}\alpha_{m,n}\log_2\left(1+\frac{s_{m,n,c}\nu_{m,n,c}}{\sum\limits_{j=1,j\neq m}^{M}\sum\limits_{i=1}^{N_j}s_{j,i,c}\nu_{j,i,c}+\sum\limits_{i=1}^{n-1}s_{m,i,c}\nu_{m,i,c}}\right).\label{eqn17}
\end{eqnarray}
\end{figure*}
It is seen in (\ref{eqn17}) that the upper bound is independent of the transmit power, and thus it will be saturated. In addition, the saturated upper bound also has nothing to do with the channel gains in the beamspace. Thus, the UEs with the same order but in different clusters may asymptotically achieve the same performance.
\end{proof}

\subsection{Performance Optimization}
From Theorem 1, the overall performance of the cellular IoT depends on the design of non-orthogonal beams in the beamspace, namely beam selection and power allocation. In this section, we jointly optimize beam selection and power allocation from the perspective of maximizing the upper bound on the weighted sum of the ergodic rates, which can be formulated as the following problem:
\begin{eqnarray}
J_1\!\!\!\!&:&\!\!\!\!\max_{\mathbf{S},\mathbf{P}}R_{\mathrm{ub}}\nonumber\\
\textrm{s.t. C1}\!\!\!\!&:&\!\!\!\!\sum\limits_{m=1}^{M}\sum\limits_{n=1}^{N_m}\sum\limits_{c=1}^{N_t}s_{m,n,c}p_{m,n,c}\leq P_{\max},\nonumber
\end{eqnarray}
where $P_{\max}$ is the maximum transmit power at the BS, $\mathbf{S}=\{s_{1,1,1},\cdots,s_{M,N_M,N_t}\}$ and $\mathbf{P}=\{p_{1,1,1},\ldots,p_{M,N_M,N_t}$\} are the beam selection and power allocation matrices, respectively. Since $s_{m,n,c},\forall m,n,c$ should be in $\{0,1\}$, $J_1$ is a mixed integer programming problem. Thus, it is difficult to obtain the optimal solutions.

Generally speaking, beam selection forms the set of the orthogonal base beams for serving the UEs, and power allocation constructs the transmit beams based on the selected base beams. Thus, the number of available orthogonal base beams determines the size of the subspace, namely the angular resolution. Inspired by this, we provide three suboptimal design algorithms from the viewpoint of the angular resolution in the beamspace.

\subsubsection{Full-Space Multiple-Beam Design}
\begin{figure}[h] \centering
\includegraphics [width=0.48\textwidth] {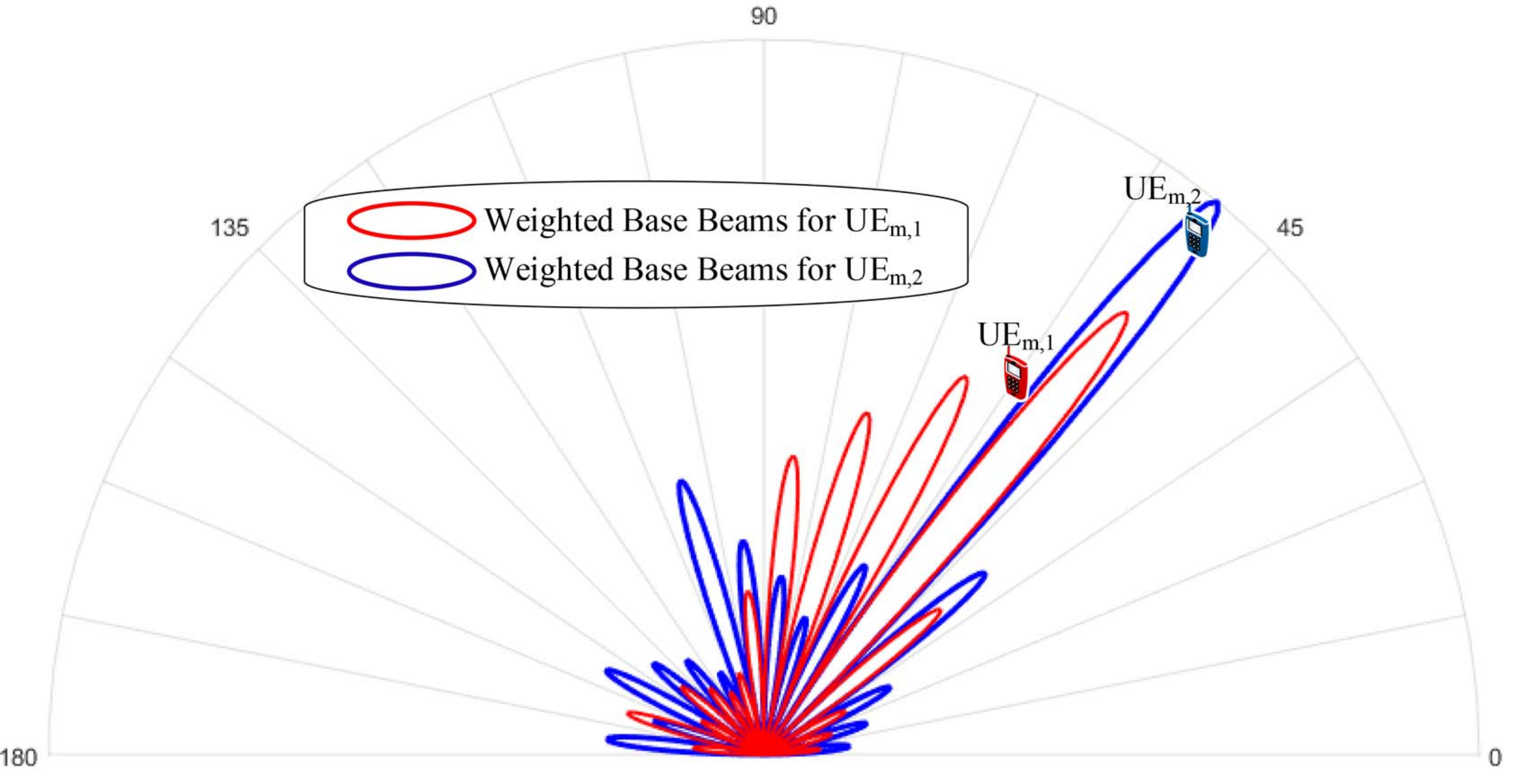}
\caption {The multiple-beam design in the whole beamspace.}
\label{fig31}
\end{figure}

First, we consider the case that the BS can select all orthogonal base beams for each UE, namely $s_{m,n,c}=1,\forall m,n,c$. As seen in Fig. \ref{fig31}, the beam for each UE can be constructed in the whole beamspace, and the transmit beams are in general non-orthogonal. In this case, the beam design is reduced to the problem of power allocation, namely determining the weighted coefficients for each UE's orthogonal base beams, which can be expressed as $J_2$ on the next page.
\begin{figure*}
\begin{eqnarray}
J_2\!\!\!\!&:&\!\!\!\!\max_{\mathbf{P}}\sum\limits_{m=1}^{M}\sum\limits_{n=1}^{N_m}\sum\limits_{c=1}^{N_t}\alpha_{m,n}\log_2\left(1+\frac{p_{m,n,c}\eta_{m,n,c}}{\sum\limits_{j=1,j\neq m}^{M}\sum\limits_{i=1}^{N_j}p_{j,i,c}\eta_{m,n,c}+\sum\limits_{i=1}^{n-1}p_{m,i,c}\eta_{m,n,c}+1}\right)\nonumber\\
\textrm{s.t. C2}\!\!\!\!&:&\!\!\!\!\sum\limits_{m=1}^{M}\sum\limits_{n=1}^{N_m}\sum\limits_{c=1}^{N_t}p_{m,n,c}\leq P_{\max}.\nonumber
\end{eqnarray}
\end{figure*}
Unfortunately, the objective function of $J_2$ is still non-concave with respect to the transmit powers, and it is difficult to design the optimal power allocation. Checking the objective function of $J_2$, it is found that it can be considered as the sum rate of $KN_t$ independent UEs, where the transmit power is $p_{m,n,c}$ and the channel gain is $\eta_{m,n,c}$ for the equivalent UE$_{m,n,c}$. Therefore, we have the following equivalent input-output relation to the UE$_{m,n,c}$:
\begin{eqnarray}
y_{m,n,c}&=&\sqrt{p_{m,n,c}\eta_{m,n,c}}x_{m,n,c}\nonumber\\
&&+\sum\limits_{j=1,j\neq m}^{M}\sum\limits_{i=1}^{N_j}\sqrt{p_{j,i,c}\eta_{m,n,c}}x_{j,i,c}\nonumber\\
&&+\sum\limits_{i=1}^{n-1}\sqrt{p_{m,i,c}\eta_{m,n,c}}x_{m,i,c}+n_{m,n,c},\label{eqn18}
\end{eqnarray}
where $x_{m,n,c}$ is the normalized Gaussian distributed desired signal for the equivalent UE$_{m,n,c}$ and $n_{m,n,c}$ is the AWGN with unit norm. Then, the SINR and the achievable rate for the equivalent UE$_{m,n,c}$ are given by
\begin{equation}
\gamma_{m,n,c}=\frac{p_{m,n,c}\eta_{m,n,c}}{\sum\limits_{j=1,j\neq m}^{M}\sum\limits_{i=1}^{N_j}p_{j,i,c}\eta_{m,n,c}+\sum\limits_{i=1}^{n-1}p_{m,i,c}\eta_{m,n,c}+1},\label{eqn19}
\end{equation}
and
\begin{equation}
R_{m,n,c}=\log_2\left(1+\gamma_{m,n,c}\right),\label{eqn20}
\end{equation}
respectively. It is well known that the minimum mean squared error (MSE) and the SINR for an arbitrary received signal have the following relation \cite{WMMSE1}, \cite{WMMSE2}:

\emph{Lemma 3}: The minimum MSE $e_{m,n,c}$ and the SINR $\gamma_{m,n,c}$ of a received signal satisfy
\begin{equation}
e_{m,n,c}^{-1}=1+\gamma_{m,n,c}.\label{eqn21}
\end{equation}
In other words, maximizing the rate is equivalent to minimizing the MSE. According to the input-output relation in (\ref{eqn18}), the MSE for the equivalent UE$_{m,n,c}$ can be easily expressed as (\ref{eqn22}),
\begin{figure*}
\begin{eqnarray}
\textrm{MSE}_{m,n,c}&=&\mathrm{E}[(v_{m,n,c}y_{m,n,c}-x_{m,n,c})(v_{m,n,c}y_{m,n,c}-x_{m,n,c})^H]\nonumber\\
&=&v_{m,n,c}\left(\eta_{m,n,c}\left(\sum\limits_{j=1,j\neq m}^{M}\sum\limits_{i=1}^{N_j}p_{j,i,c}+\sum\limits_{i=1}^{n}p_{m,i,c}\right)+1\right)v_{m,n,c}^H\nonumber\\
&&-\sqrt{\eta_{m,n,c}p_{m,n,c}}\left(v_{m,n,c}+v_{m,n,c}^H\right)+1\nonumber\\
&=&\left(v_{m,n,c}-\sqrt{\eta_{m,n,c}p_{m,n,c}}\Phi_{m,n,c}^{-1}\right)\Phi_{m,n,c}\left(v_{m,n,c}-\sqrt{\eta_{m,n,c}p_{m,n,c}}\Phi_{m,n,c}^{-1}\right)^H\nonumber\\
&&-\eta_{m,n,c}p_{m,n,c}\Phi_{m,n,c}^{-1}+1,\label{eqn22}
\end{eqnarray}
\end{figure*}
where $v_{m,n,c}$ denotes the receiver, and $\Phi_{m,n,c}=\eta_{m,n,c}\left(\sum\limits_{j=1,j\neq m}^{M}\sum\limits_{i=1}^{N_j}p_{j,i,c}+\sum\limits_{i=1}^{n}p_{m,i,c}\right)+1$ is the power of the equivalent received signal of the UE$_{m,n,c}$. It is clear that $\textrm{MSE}_{m,n,c}$ is minimized if and only if $v_{m,n,c}=\sqrt{\eta_{m,n,c}p_{m,n,c}}\Phi_{m,n,c}^{-1}$, namely adopting the MMSE receiver.

Thus, according to Lemma 3, the objective function of $J_2$ can be transformed as
\begin{equation}
\min_{\mathbf{P}}\sum\limits_{m=1}^{M}\sum\limits_{n=1}^{N_m}\sum\limits_{c=1}^{N_t}\alpha_{m,n}(\log_2(e_{m,n,c})).\label{eqn23}
\end{equation}
However, the objective function in (\ref{eqn23}) is still not convex. Note that (\ref{eqn23}) aims to minimize a function of minimum MSE, which is equivalent to minimizing a function of MSE for a given MMSE receiver. In other words, the optimization objective in (\ref{eqn23}) can be transformed as
\begin{equation}
\min_{\mathbf{P},\mathbf{v}}\sum\limits_{m=1}^{M}\sum\limits_{n=1}^{N_m}\sum\limits_{c=1}^{N_t}\alpha_{m,n}(\log_2(\mathrm{MSE}_{m,n,c})),\label{eqn24}
\end{equation}
where $\mathbf{v}=\{v_{1,1,1},\cdots,v_{M,N_M,N_t}\}$ is the collection of the receivers. The sum of logarithmic functions hinders us to further solve this problem. Similar to \cite{WMMSE1} and \cite{WMMSE2}, we can replace the logarithmic function with the following term
\begin{eqnarray}
\min_{\mathbf{P},\mathbf{v},\bm{\beta}}\sum\limits_{m=1}^{M}\sum\limits_{n=1}^{N_m}\sum\limits_{c=1}^{N_t}\alpha_{m,n}(\beta_{m,n,c}\textrm{MSE}_{m,n,c}-\log_2(\beta_{m,n,c})),\label{eqn25}
\end{eqnarray}
where $\bm{\beta}=\{\beta_{1,1,1},\cdots,\beta_{M,N_M,N_t}\}$ is the collection of auxiliary variables. Note that only when $\beta_{m,n,c}=\textrm{MSE}_{m,n,c}^{-1}$, the objective function in (\ref{eqn25}) can achieve its minimum value. Under such a condition, the optimization objectives (\ref{eqn24}) and (\ref{eqn25}) are equivalent.

According to the definition of $\mathrm{MSE}_{m,n,c}$ in (\ref{eqn22}), it is known that (\ref{eqn25}) is not a jointly convex function of $\mathbf{P}$, $\mathbf{v}$, and $\bm{\beta}$, but is a convex function with respect to each optimization variable. Thus, we can adopt the sequential iteration optimization method to solve the problem. Specifically, we optimize one variable by fixing the others, and the variables are iteratively optimized until they approach a stationary point. First, for the variable $\mathbf{P}$, by combining the objective function (\ref{eqn25}) and the constraint condition C2, we get the Lagrange function as% (\ref{eqn26}) at the top of the next page,
\begin{eqnarray}
\mathcal{L}_1(\mathbf{P})\!\!\!\!\!&=&\!\!\!\!\!\!\sum\limits_{m=1}^{M}\sum\limits_{n=1}^{N_m}\sum\limits_{c=1}^{N_t}\alpha_{m,n}(\beta_{m,n,c}\mathrm{MSE}_{m,n,c}-\log_2(\beta_{m,n,c}))\nonumber\\
&&\!\!\!\!\!\!+\mu\left(\sum\limits_{m=1}^{M}\sum\limits_{n=1}^{N_m}\sum\limits_{c=1}^{N_t}p_{m,n,c}-P_{\max}\right),\label{eqn26}
\end{eqnarray}
%\begin{figure*}
%\begin{eqnarray}
%\mathcal{L}_1(\mathbf{P})&=&\sum\limits_{m=1}^{M}\sum\limits_{n=1}^{N_m}\sum\limits_{c=1}^{N_t}\alpha_{m,n}(\beta_{m,n,c}\mathrm{MSE}_{m,n,c}-\log_2(\beta_{m,n,c}))+\mu\left(\sum\limits_{m=1}^{M}\sum\limits_{n=1}^{N_m}\sum\limits_{c=1}^{N_t}p_{m,n,c}-P_{\max}\right),\label{eqn26}
%\end{eqnarray}
%\end{figure*}
where $\mu\geq0$ is the Lagrange multiplier of C2. By leveraging the KKT conditions, we obtain Eq. (\ref{eqn27}) at the top of the next page.
\begin{figure*}
\begin{equation}
p_{m,n,c}=\left(\frac{\alpha_{m,n}\beta_{m,n,c}v_{m,n,c}\sqrt{\eta_{m,n,c}}}{\sum\limits_{j=1,j\neq m}^{M}\sum\limits_{i=1}^{N_m}\alpha_{j,i}\beta_{j,i,c}v_{j,i,c}^2\eta_{j,i,c}+\sum\limits_{i=1}^{n}\alpha_{m,i}\beta_{m,i,c}v_{m,i,c}^2\eta_{m,i,c}+\mu}\right)^2.\label{eqn27}
\end{equation}
\end{figure*}
Then, the intermediate variables $\mathbf{v}$ and $\bm{\beta}$ capturing the performance of the UEs in the previous iteration can be obtained according to their definitions. Moreover, the Lagrange multiplier $\mu$ can be updated by the iterative gradient method. In the $(t+1)$th iteration, the $\mu$ can be updated as
\begin{equation}
\mu(t+1)=\left[\mu(t)+\Delta_{\mu}\left(\sum\limits_{m=1}^{M}\sum\limits_{n=1}^{N_m}\sum\limits_{c=1}^{N_t}p_{m,n,c}-P_{\max}\right)\right]^+,\label{eqn28}
\end{equation}
where $\Delta_{\mu}>0$ is an iteration step size. In summary, the full-space multiple-beam design algorithm can be described as

\begin{algorithm}
\SetAlgoNoLine
\caption{Full-Space Multiple-Beam Design}
\label{alg1}
\KwIn{$\alpha_{m,n}, P_{\max}, \eta_{m,n,c}$}
\KwOut{$\mathbf{P}$}
Initialize the parameter: $p_{m,n,c}=\frac{P_{\max}}{KN_t},\forall m,n,c$\;
\SetKwRepeat{doWhile}{do}{while}
\doWhile{$\mathbf{P}$ is not converged}{
$\mu=1,v_{m,n,c}=\sqrt{\eta_{m,n,c}p_{m,n,c}}\Phi_{m,n,c}^{-1},$
$\beta_{m,n,c}=\textrm{MSE}_{m,n,c}^{-1},\forall m,n,c$\;
\doWhile{$\mu$ is not converged}{
Update $p_{m,n,c}$ according to (\ref{eqn27}), $\forall m,n,c$,\\
Update $\mu$ by the gradient method in (\ref{eqn28})\;
}}
\end{algorithm}

In this algorithm, we construct the transmit beam in the whole beamspace, and thus there are $KN_t$ optimization variables for $\mathbf{P}$ in total. In fact, since the UEs are partitioned into clusters according to the AoD information, the transmit beam for an arbitrary UE is mainly determined by a finite number of base beams with high correlation in the beamspace. Thus, the required number of optimization variables for $\mathbf{P}$ can be effectively reduced.

\subsubsection{Partial-Space Multiple-Beam Design}
\begin{figure}[h] \centering
\includegraphics [width=0.48\textwidth] {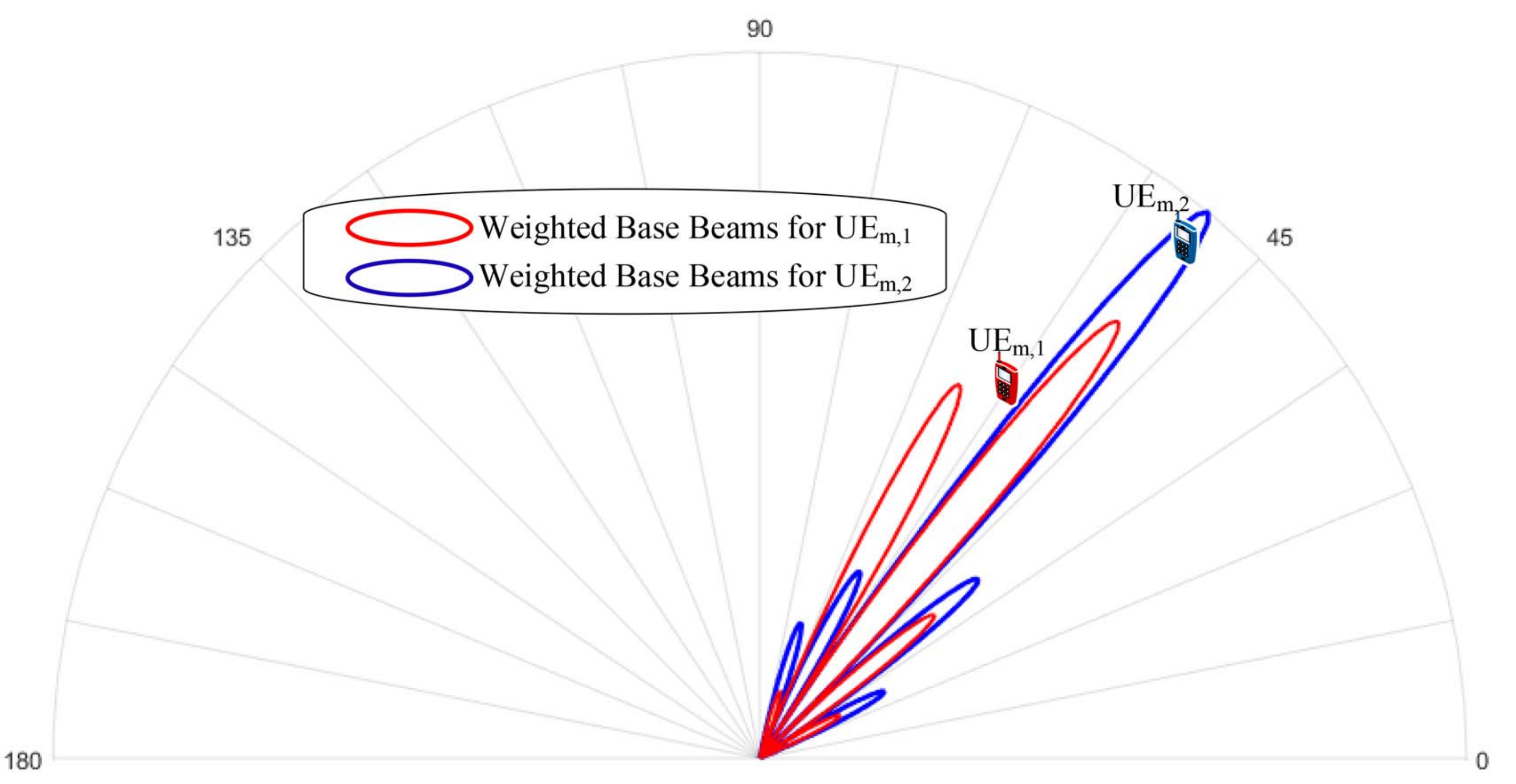}
\caption {The multiple-beam design in the partial beamspace.}
\label{fig32}
\end{figure}

In this section, we design the transmit beam for each UE with a few base beams, namely in the partial beamspace, cf. Fig. \ref{fig32}. Note that in the scenario of massive access for the cellular IoT, the system performance is mainly limited by the co-channel interference. In order to effectively reduce the co-channel interference, we enforce the subspaces between the clusters orthogonal to each other. In specific, the $N_t$ base beams are allocated to the $M$ clusters, and a base beam can only serve the UEs in a cluster, namely $\sum_{m=1}^{M}s_{m,n,c}=1$. As a result, we have $\sum\limits_{j=1,j\neq m}^{M}\sum\limits_{i=1}^{N_j}s_{j,i,c}p_{j,i,c}\eta_{m,n,c}=0, \forall j\neq m$. In this case, the beam design can be formulated as the problem $J_3$ at the top of this page.
\begin{figure*}
\begin{eqnarray}
J_3\!\!\!\!&:&\!\!\!\!\max_{\mathbf{S},\mathbf{P}}\sum\limits_{m=1}^{M}\sum\limits_{n=1}^{N_m}\sum\limits_{c=1}^{N_t}\alpha_{m,n}\log_2\left(1+\frac{s_{m,n,c}p_{m,n,c}\eta_{m,n,c}}{\sum\limits_{i=1}^{n-1}s_{m,i,c}p_{m,i,c}\eta_{m,n,c}+1}\right)\nonumber\\
\textrm{s.t. C1},\textrm{C3}\!\!\!\!&:&\!\!\!\!\sum\limits_{m=1}^{M}s_{m,n,c}=1, \forall n,c.\nonumber
\end{eqnarray}
\end{figure*}
As mentioned above, due to the constraint of $s_{m,n,c}\in\{0,1\}$, $J_3$ is also a mixed integer programming problem. To solve this problem, we partition it into two subproblems, one for beam selection, and the other for power allocation.

We first address the problem of beam selection. It is intuitive that the optimal beam selection can be realized by the exhaustive searching. However, since there might be a massive number of clusters and base beams, the computational complexity of the exhaustive searching is prohibitive. Hence, we propose a low-complexity beam selection method according to the characteristics of massive access in the beamspace. Checking the objective function of $J_3$, it is found that since a base beam is only distributed to one cluster exclusively, there is no inter-cluster interference. In other words, the clusters are independent of each other over an arbitrary base beam. Meanwhile, there is no interrelation among the base beams. Thus, we can allocate the base beams one by one. For a certain base beam, in order to maximize the weighted sum rate, it is better to allocate it to the cluster with the maximum weighted sum rate over such a base beam. Mathematically, the beam selection method can be expressed as
\begin{equation}
m^{\star}\!=\!\arg\max\limits_{m=1,\cdots,M}\sum\limits_{n=1}^{N_m}\alpha_{m,n}\log_2\!\left(1+\frac{p_{m,n,c}\eta_{m,n,c}}{\sum\limits_{i=1}^{n-1}p_{m,i,c}\eta_{m,n,c}+1}\right)\!.\label{eqn29}
\end{equation}
In other words, the $c$th base beam is allocated to the $m^{\star}$th cluster.

Then, given the beam selection result, we allocate the power for constructing the transmit beam of each UE. Fortunately, based on the selected base beams, power allocation is similar to the optimization problem $J_2$. Therefore, with the same method, if the $c$th base beam is allocated to the $m$th cluster, the transmit power $p_{m,n,c}$ is given by
\begin{equation}
p_{m,n,c}=\left(\frac{\alpha_{m,n,c}\beta_{m,n,c}^{'}v_{m,n,c}^{'}\sqrt{\eta_{m,n,c}}}{\sum\limits_{i=1}^{n}\alpha_{m,i,c}\beta_{m,i,c}^{'}(v_{m,i,c}^{'})^2\eta_{m,i,c}+\mu^{'}}\right)^2,\label{eqn30}
\end{equation}
where
\begin{equation}
v_{m,n,c}^{'}=\sqrt{\eta_{m,n,c}p_{m,n,c}}\left(\eta_{m,n,c}\sum\limits_{i=1}^{n}p_{m,i,c}+1\right)^{-1},\label{eqn31}
\end{equation}
\begin{eqnarray}
\beta_{m,n,c}^{'}\!\!\!\!\!&=&\!\!\!\!\!\Bigg(v_{m,n,c}^{'}\left(\eta_{m,n,c}\sum\limits_{i=1}^{n}p_{m,i,c}+1\right)(v_{m,n,c}^{'})^H\nonumber\\
&&\!\!\!\!\!\!-\sqrt{\eta_{m,n,c}p_{m,n,c}}\left(v_{m,n,c}^{'}+(v_{m,n,c}^{'})^H\right)+1\Bigg)^{-1},\nonumber\\\label{eqn32}
\end{eqnarray}
and the Lagrange multiplier $\mu^{'}$ can be updated as
\begin{equation}
\mu^{'}(t+1)\!=\!\left[\mu^{'}(t)\!+\!\Delta_{\mu'}\!\left(\sum\limits_{m=1}^{M}\sum\limits_{n=1}^{N_m}\sum\limits_{c=1}^{N_t}s_{m,n,c}p_{m,n,c}\!-\!P_{\max}\right)\!\right]^+,\label{eqn33}
\end{equation}
where $\Delta_{\mu'}>0$ is an iteration step size. Thus, the partial-space multiple-beam design algorithm can be summarized as Algorithm 2 on the next page.
\begin{algorithm}
\SetAlgoNoLine
\caption{Partial-Space Multiple-Beam Design}
\label{alg2}
\KwIn{$\alpha_{m,n}, P_{\max}, \eta_{m,n,c}$}
\KwOut{$\mathbf{P},\mathbf{S}$}
Initialize the parameter: $s_{m,n,c}=0$, $p_{m,n,c}=\frac{P_{\max}}{N_mN_t},\forall m,n,c$\;
\For{$c=1:N_t$}
{
Choose $m^{\star}$ and let $s_{m^{\star},n,c}=1$ according to (\ref{eqn29}), $\forall n$\;
Let $p_{j,n,c}=0,\forall j\neq m^{\star},n$\;
}
\SetKwRepeat{doWhile}{do}{while}
\doWhile{$\mathbf{P}$ is not converged}{
$\mu'=1,v_{m,n,c}'$ and $\beta_{m,n,c}'$ are updated according to (\ref{eqn31}) and (\ref{eqn32}), respectively, $\forall m,n,c$\;
\doWhile{$\mu'$ is not converged}{
Update $p_{m,n,c}$ according to (\ref{eqn30}), $\forall m,n,c$,\\
Update $\mu^{'}$ by the gradient method in (\ref{eqn33})\;
}}
\end{algorithm}

In this algorithm, we construct a transmit beam for each UE. Thus, we need to have $K$ transmit beams in total. In the scenario of massive access for the cellular IoT, the required number of transmit beams might be very large, resulting in a high computational complexity. In fact, since we perform user clustering according to the AoD information, the UEs in a cluster may have small differences in the AoD. Especially in the case of a large-scale antenna array at the BS, the AoDs of the UEs in a cluster are nearly the same. In this context, we can construct the same transmit beam for a cluster, and hence the required number of transmit beams and the corresponding computational complexity can be reduced significantly.

\subsubsection{Partial-Space Single-Beam Design}
\begin{figure}[h] \centering
\includegraphics [width=0.48\textwidth] {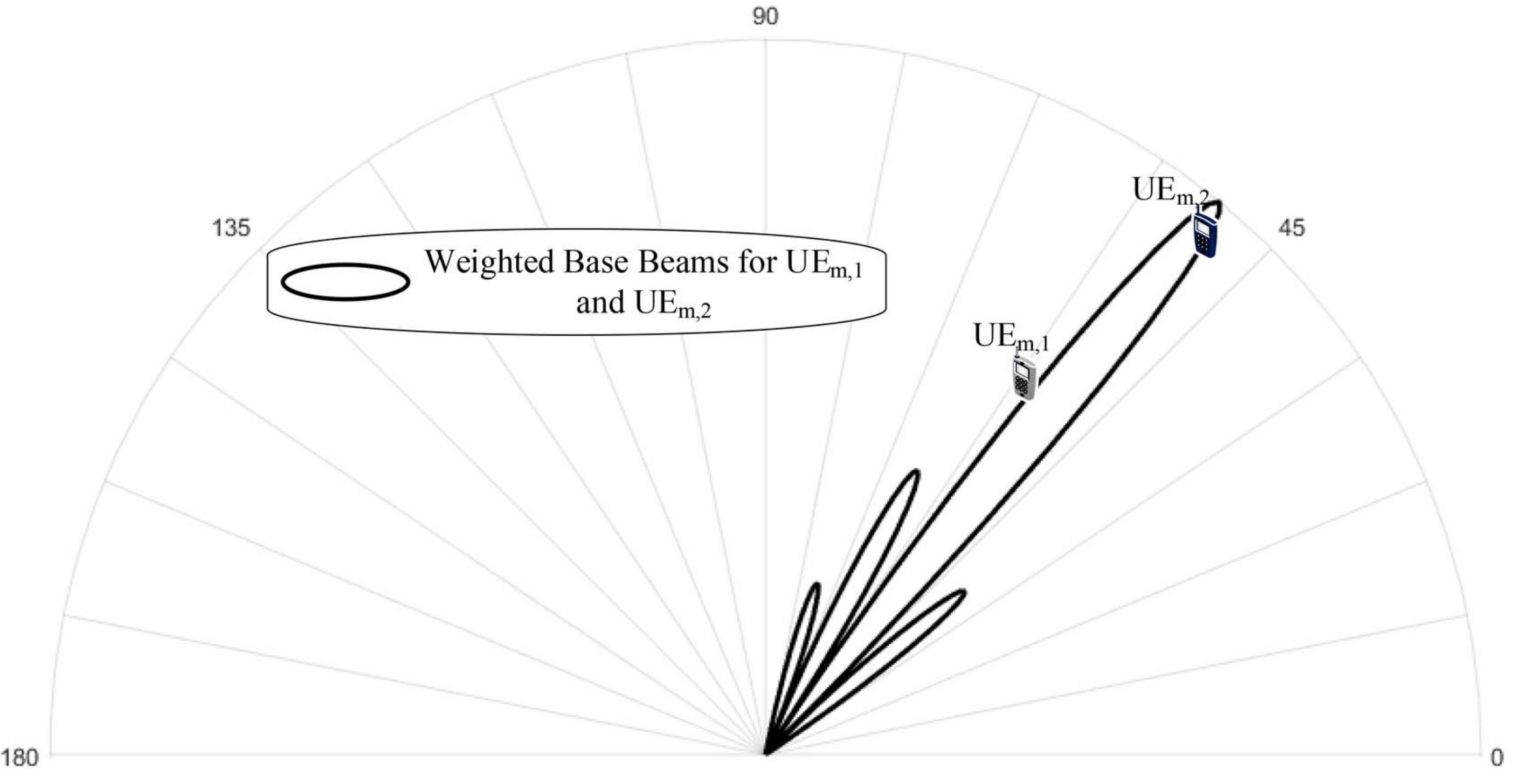}
\caption {The single-beam design in the partial beamspace.}
\label{fig33}
\end{figure}

As mentioned above, considering the high correlation among the UEs in a cluster, we can design only one transmit beam for a cluster, which can reduce the computation complexity but also decreases the degrees of freedom. Meanwhile, to mitigate the inter-cluster interference, the subspaces, namely the base beams, for designing the transmit beams are orthogonal of each other. In general, as shown in Fig. \ref{fig33}, the total $N_t$ base beams are divided into $M$ sets, and each set of base beams is used to construct a transmit beam for a specific cluster.

It is assumed that the transmit beam for the $m$th cluster is $\mathbf{w}_m$, then it can be constructed as
\begin{equation}
\mathbf{w}_m=\sum\limits_{c\in\mathcal{B}_m}\sqrt{p_c}\mathbf{u}_c,
\end{equation}
where $\mathcal{B}_m$ is the index collection of the selected base beams for the $m$th cluster, and $p_c$ is the total transmit power over the $c$th base beam. In order to guarantee that the UEs in a cluster share the same transmit beam, $\mathbf{w}_{m}$ and $\mathbf{w}_{m,n}$ should be aligned, $\forall n$. Equivalently, the transmit power of the UE$_{m,n}$ over the $c$th base beam $p_{m,n,c}$ should satisfy the following condition
\begin{eqnarray}
p_{m,n,c}=\iota_{m,n}p_c,\label{eqn35}
\end{eqnarray}
where $0\leq\iota_{m,n}\leq1$ is the power allocation factor of the UE$_{m,n}$ with the constraint $\sum\limits_{n=1}^{N_m}\iota_{m,n}=1$. Then, the relation between $\mathbf{w}_{m}$ and $\mathbf{w}_{m,n}$ can be expressed as
\begin{equation}
\mathbf{w}_{m,n}=\sqrt{\iota_{m,n}}\mathbf{w}_m=\sqrt{\iota_{m,n}}\sum\limits_{c\in\mathcal{B}_m}\sqrt{p_c}\mathbf{u}_c=\sum\limits_{c\in\mathcal{B}_m}\sqrt{p_{m,n,c}}\mathbf{u}_c,\label{eqn34}
\end{equation}
which satisfies the power equality in (\ref{eqn35}), and $\mathbf{w}_{m,n}$ is also aligned with $\mathbf{w}_{m}$.

Thus, the partial-space single-beam design can be formulated as an optimization problem $J_4$ at the top of the next page,
\begin{figure*}
\begin{eqnarray}
J_4\!\!\!\!&:&\!\!\!\!\max_{\mathbf{S},\mathbf{p}_c,\bm{\iota}}\sum\limits_{m=1}^{M}\sum\limits_{n=1}^{N_m}\sum\limits_{c=1}^{N_t}\alpha_{m,n}\log_2\left(1+\frac{s_{m,n,c}p_c\iota_{m,n}\eta_{m,n,c}}{s_{m,n,c}p_c\eta_{m,n,c}\sum\limits_{i=1}^{n-1}\iota_{m,i}+1}\right)\nonumber\\
\textrm{s.t. C3, C4}\!\!\!\!&:&\!\!\!\!\sum\limits_{c=1}^{N_t}p_{c}\leq P_{\max},\nonumber\\
\textrm{C5}\!\!\!\!&:&\!\!\!\!\sum\limits_{n=1}^{N_m}\iota_{m,n}=1, \forall m,\nonumber
\end{eqnarray}
\end{figure*}
where $\mathbf{p}_c=\{p_1,\ldots,p_c\}$ and $\bm{\iota}=\{\iota_{1,1},\cdots,\iota_{M,N_M}\}$. Similarly, $J_4$ is a mixed integer programming problem, and we solve it through beam selection and power allocation separately. For beam selection, we can adopt the same method as (\ref{eqn29}) in the last algorithm.

Then, we design the power allocation method as before. First, given the beam selection result, the Lagrange function of $J_4$ can be written as
\begin{eqnarray}
\mathcal{L}_2(\mathbf{p}_c,\bm{\iota})\!\!\!\!\!&=&\!\!\!\!\!\!\sum\limits_{m=1}^{M}\sum\limits_{n=1}^{N_m}\sum\limits_{c=1}^{N_t}\alpha_{m,n}(\beta_{m,n,c}''\mathrm{MSE}_{m,n,c}''-\log_2(\beta_{m,n,c}''))\nonumber\\
&&\!\!\!\!\!\!+\mu''\left(\sum\limits_{c=1}^{N_t}p_c-P_{\max}\right)\nonumber\\
&&\!\!\!\!\!\!+\sum\limits_{m=1}^{M}\omega_m\left(\sum\limits_{n=1}^{N_m}\iota_{m,n}-1\right),\label{eqn36}
\end{eqnarray}
where
\begin{eqnarray}
\!\!\!\!\!\!\!\mathrm{MSE}_{m,n,c}''\!\!\!\!\!&=&\!\!\!\!\!v_{m,n,c}''\left(\eta_{m,n,c}p_c\sum\limits_{i=1}^{n}\iota_{m,i}+1\right)(v_{m,n,c}'')^H\nonumber\\
&&\!\!\!\!\!-\sqrt{\eta_{m,n,c}p_c\iota_{m,n}}\left(v_{m,n,c}''+(v_{m,n,c}'')^H\right)+1,\label{eqn37}
\end{eqnarray}
\begin{equation}
v_{m,n,c}''=\sqrt{\eta_{m,n,c}p_c\iota_{m,n}}\left(\eta_{m,n,c}p_c\sum\limits_{i=1}^{n}\iota_{m,i}+1\right)^{-1},\label{eqn38}
\end{equation}
and $\beta_{m,n,c}''=(\mathrm{MSE}_{m,n,c}'')^{-1}$. Moreover, $\mu''\geq0$ and $\omega_m\geq0$ are the Lagrange multipliers of C4 and C5, respectively. Then, by using the KKT conditions, we have
\begin{equation}
p_c=\left(\frac{\sum\limits_{i=1}^{N_{m^{\star}}}\alpha_{{m^{\star}},i}\beta_{{m^{\star}},i,c}^{''}v_{{m^{\star}},i,c}^{''}\sqrt{\eta_{{m^{\star}},i,c}\iota_{{m^{\star}},i} }}{\mu''+\sum\limits_{i=1}^{N_{m^{\star}}}\alpha_{{m^{\star}},i}\beta_{{m^{\star}},i,c}^{''}(v_{{m^{\star}},i,c}^{''})^2\eta_{{m^{\star}},i,c}\sum\limits_{q=1}^{i}\iota_{{m^{\star}},q}}\right)^2.\label{eqn39}
\end{equation}
and
\begin{equation}
\iota_{m,n}=\left(\frac{\sum\limits_{c=1}^{N_t}s_{m,n,s}\alpha_{m,n}\beta_{m,n,c}^{''}v_{m,n,c}^{''}\sqrt{\eta_{m,n,c}p_c}}{\sum\limits_{c=1}^{N_t}\sum\limits_{i=n}^{N_m}s_{m,n,s}\alpha_{m,i}\beta_{m,i,c}^{''}(v_{m,i,c}^{''})^2\eta_{m,i,c}p_c+\omega_n}\right)^2,\label{eqn40}
\end{equation}
where $m^{\star}$ is the index of the cluster which uses the $c$th base beam according to (\ref{eqn28}). Moreover, $\mu''$ and $\omega_m$ can be updated by the following gradient methods
\begin{equation}
\mu^{''}(t+1)=\left[\mu^{''}(t)+\Delta_{\mu^{''}}\left(\sum\limits_{c=1}^{N_t}p_{c}-P_{\max}\right)\right]^+,\label{eqn41}
\end{equation}
and
\begin{equation}
\omega_m(t+1)=\left[\omega_m(t)+\Delta_{\omega_m}\left(\sum\limits_{n=1}^{N_m}\iota_{m,n}-1\right)\right]^+,\forall m,\label{eqn42}
\end{equation}
where $\Delta_{\mu^{''}}>0$ and $\Delta_{\omega_m}>0$ are iteration step sizes. Thus, the partial-space single beam design algorithm for massive access can be summarized as
\begin{algorithm}
\SetAlgoNoLine
\caption{Partial-Space Single-Beam Design}
\label{alg3}
\KwIn{$\alpha_{m,n}, P_{\max}, \eta_{m,n,c}$}
\KwOut{$\mathbf{S},\mathbf{P}$}
Initialize the parameter: $s_{m,n,c}=0, p_c=\frac{P_{\max}}{N_t}, \forall c,$ $\iota_{m,n}=\frac{1}{N_m},\forall m,n$\;
\For{$c=1:N_t$}
{
Choose $m^{\star}$ and let $s_{m^{\star},n,c}=1$ according to (\ref{eqn29}), $\forall n$\;
}
\SetKwRepeat{doWhile}{do}{while}
\doWhile{$\mathbf{p}_c$ is not converged}{
$\mu''=1,\omega_m=1,\forall m$,\\
$v_{m,n,c}^{''}=\sqrt{\eta_{m,n,c}p_c\iota_{m,n}}\left(\eta_{m,n,c}p_c\sum\limits_{i=1}^{n}\iota_{m,i}+1\right)^{-1},$
$\beta_{m,n,c}''=(\mathrm{MSE}_{m,n,c}'')^{-1},\forall m,n,c$\;
\doWhile{$\mu''$ or $\omega_m$ is not converged}{
Update $p_c$ according to (\ref{eqn39}), $\forall c$,\\
Update $\iota_{m,n}$ according to (\ref{eqn40}), and let $p_{m,n,c}=\iota_{m,n}p_c$, $\forall m,n$,\\
Update $\mu''$ and $\omega_m$ by the gradient methods in (\ref{eqn41}) and (\ref{eqn42}), respectively\;
}}
\end{algorithm}

Up to now, we have already presented three algorithms to design non-orthogonal transmit beams for massive access in the cellular IoT. The computation complexity of three algorithms are as follows: 1) Algorithm 1 calculates a power for each UE in each beam, so the computation complexity is $O(KN_t)$. 2) For Algorithm 2, the computation complexity of beam selection can be ignored because this part is executed only once and we only focus on the computation complexity in the iteration. The computation complexity of power allocation is $O(\bar{N}N_t)$, where $\bar{N}$ is a mean value of each cluster's UE number $N_m$. 3) The numbers of parameters in Algorithm 3 are $N_t$ ($p_c$) and $K$ ($\iota_{m,n}$). Therefore, the computation complexity can be approximated as $O(N_t+K)$.

As analyzed above, for the three algorithms, the degrees of freedom for beam selection are decreased but the number of optimization variables is also reduced in sequence. Therefore, it is possible to choose a proper beam design algorithm according to the requirements of system performance and computational complexity.

\section{Simulation Results}
To validate the effectiveness of the proposed algorithms for non-orthogonal beamspace multiple access in the cellular IoT with massive connections, we carry out extensive numerical simulations in such a scenario: $N_t=64$, $K=60$, and $P_{\max}=10$ dB. The UEs are uniformly distributed in a circle with the BS as the centre and of the radius 50m. Thus, the number of clusters is dynamically changed as the UEs move. We adopt a typical urban (TU) channel model for the cellular IoT according to the document of 3GPP TR 45.820 \cite{3GPP}. For ease of notation, we use SNR (in dB) to represent the term $10\log_{10}P_{\max}$. Specifically, the transmit
power is measured with respect to the power of noise. For example, the noise power of UE is $\sigma^2$, then the maximum transmit power is $10\sigma^2$ with $P_{\max}=10$ dB. All curves are obtained by averaging over 1000 channel realizations.

\begin{figure}[h] \centering
\includegraphics [width=0.45\textwidth] {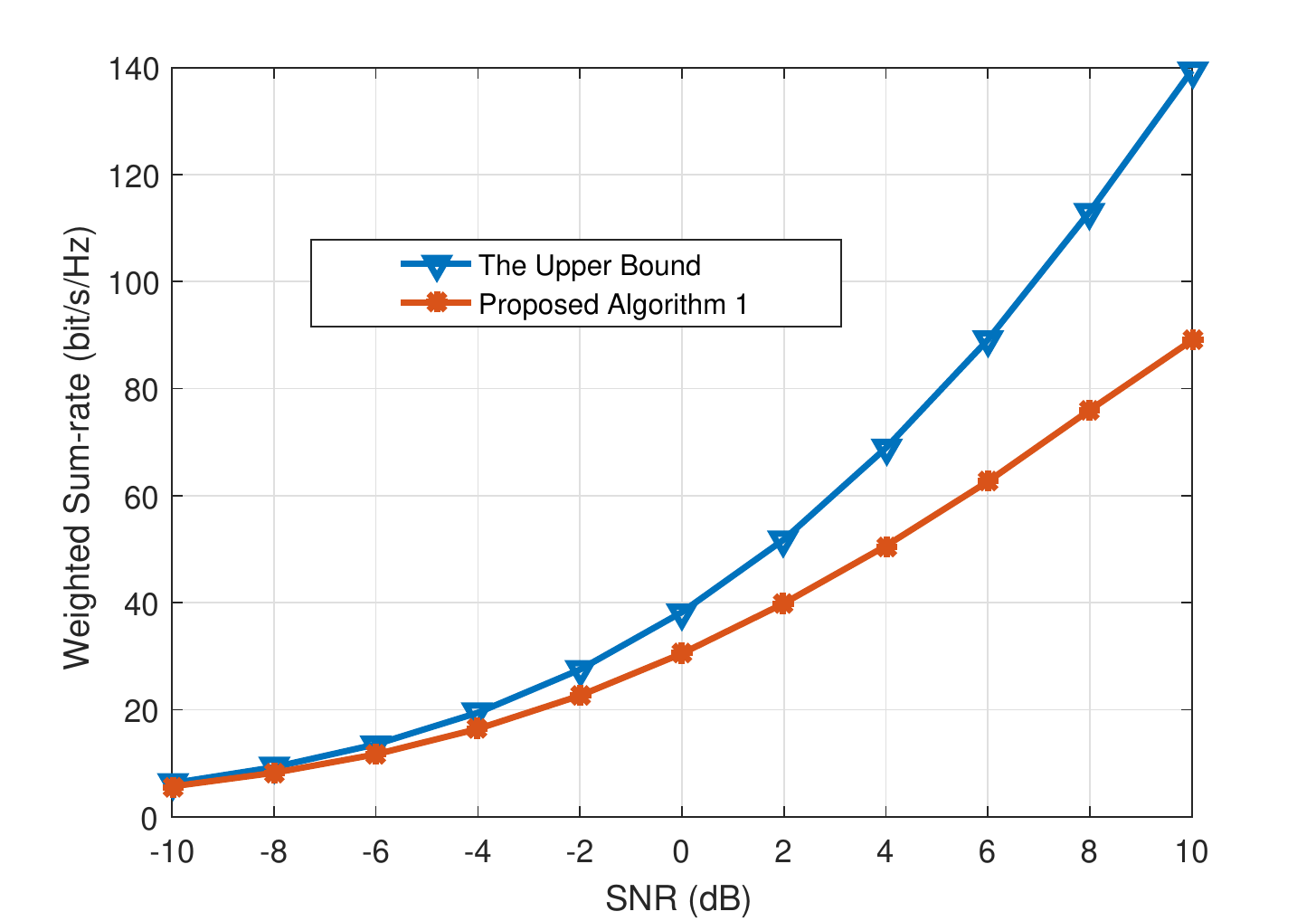}
\caption {The gap between the exact sum rate and the derived upper bound of proposed algorithm 1.}
\label{simub}
\end{figure}

First, we show the gap between the exact sum rate and the derived upper bound of the proposed algorithm 1. As seen in Fig. \ref{simub}, the gap between the exact sum rate and the upper bound is negligible when the SNR is less than 2 dB, which means that the derived upper bound provides an appropriate method to evaluate the weighted sum rate. As SNR increases, the gap becomes large, but the exact sum rate and the upper bound have the same trend.

\begin{figure}[h] \centering
\includegraphics [width=0.45\textwidth] {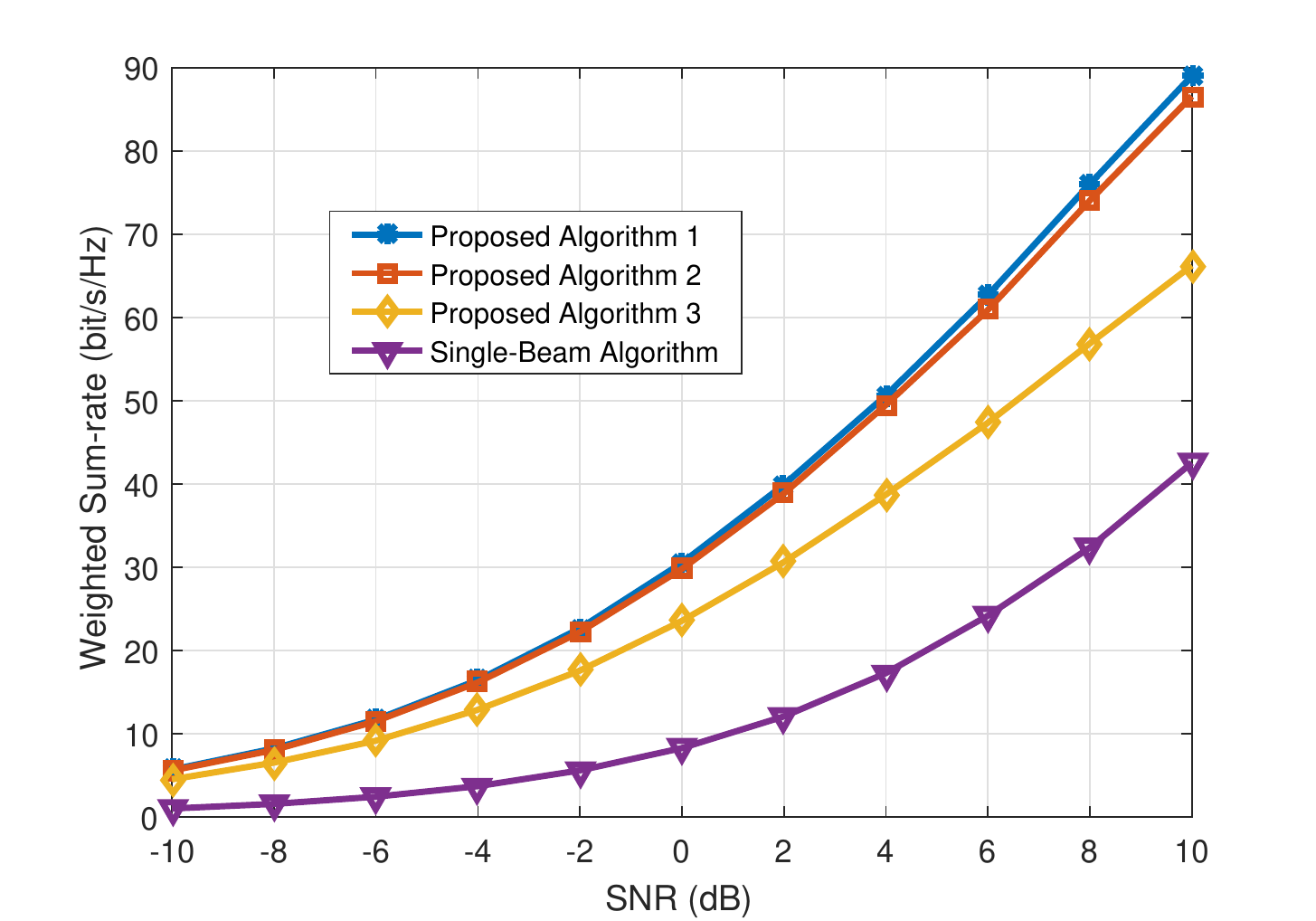}
\caption {Performance comparison of different beam design algorithms.}
\label{sim1}
\end{figure}

Then, we compare the performance of the three proposed algorithms and the traditional single-beam algorithm in Fig. \ref{sim1}. Specifically, the traditional single-beam algorithm (namely beamspace MF) uses the base beam specified by the AoD information as the transmit beam for a given cluster. It is seen that the three proposed multiple-beam combination algorithms perform much better than the single-beam algorithm, and the performance gain enlarges as the SNR increases. This is because the proposed algorithms have more degrees of freedom to design the transmit beams, then can mitigate the co-channel interference to a large extend. Therefore, the proposed algorithms have a strong capability to support massive access over limited radio spectrum. For the three proposed algorithms, the second one can nearly achieve the same performance as the first one in the whole SNR region. In other words, it is enough to construct the transmit beam based on a few base beams with high correlation. Since the second one can obtain a balance between system performance and computational complexity, we take it as a typical non-orthogonal beamspace multiple access algorithm for performance comparison with the other multiple access technique in the following.

\begin{figure}[h] \centering
\includegraphics [width=0.45\textwidth] {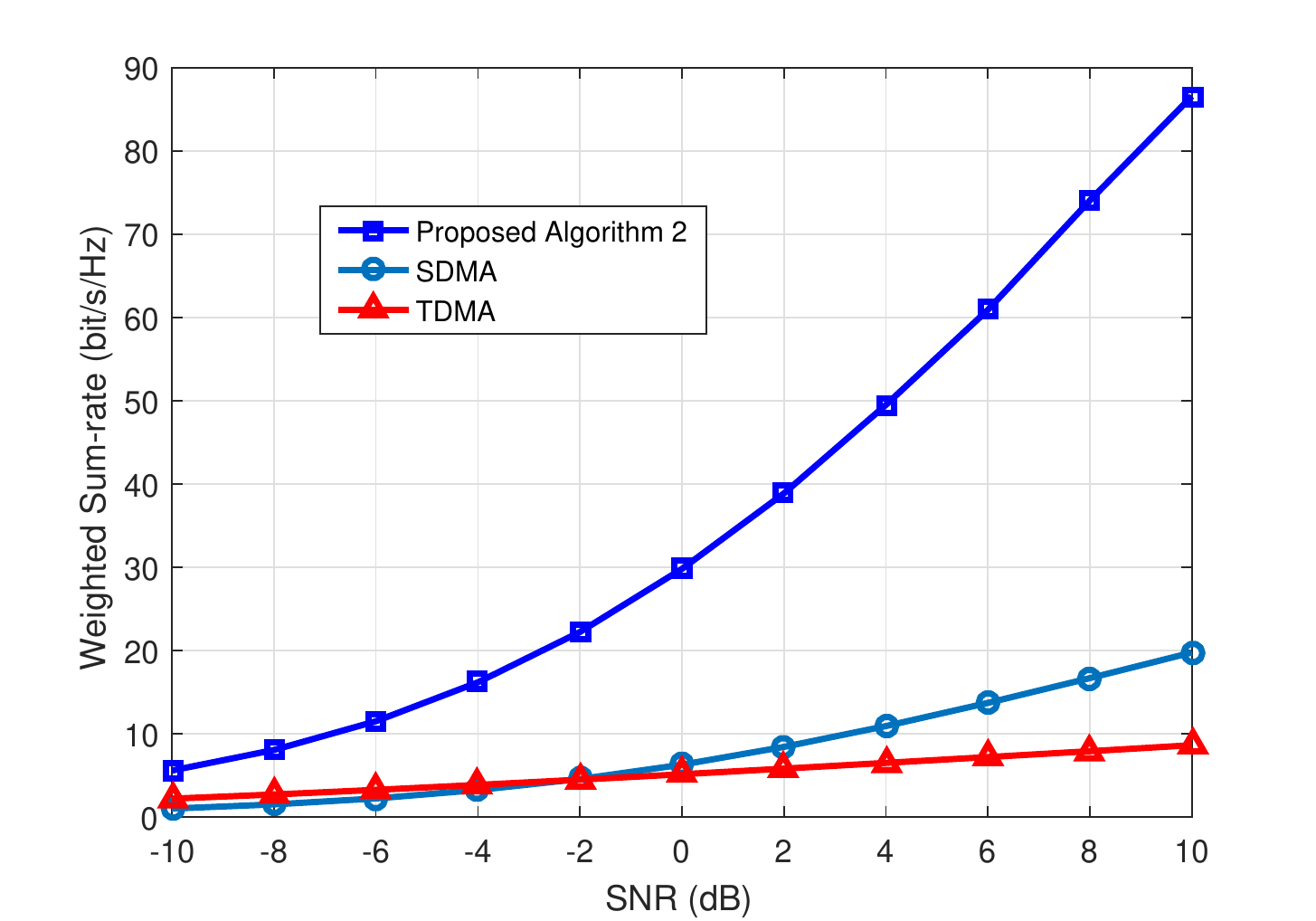}
\caption {Performance comparison of the NOMA and OMA schemes.}
\label{sim2}
\end{figure}

Fig. \ref{sim2} exhibits the performance advantage of the proposed non-orthogonal beamspace multiple access scheme (algorithm 2) over two traditional OMA schemes, including space division multiple accse (SDMA) and time division multiple access (TDMA). To be specific, the TDMA scheme divides a time slot into $K$ sub-slots and each UE occupies a sub-slot exclusively, while the SDMA scheme serves all UEs at the same time. Both TDMA and SDMA select the optimal base beam as the transmit beam for each UE, which lead to a fair comparison because all schemes are designed according to the same beam domain channel information. It is seen that the proposed algorithm 2 has an obvious performance gain over the SDMA. This is because the proposed algorithm 2 has more degrees of freedom to design the transmit beam and the SIC can further mitigate the interference. Comparing with TDMA, the proposed algorithm 2 effectively exploits the spatial multiplexing capability offered by the multiple-antenna BS, and thus can significantly improve the sum rate. Although there is high co-channel interference in the high SNR region, the proposed algorithm 2 is able to mitigate the interference by using beamspace beamforming. Hence, the performance gain becomes large as the SNR increases.

\begin{figure}[h] \centering
\includegraphics [width=0.45\textwidth] {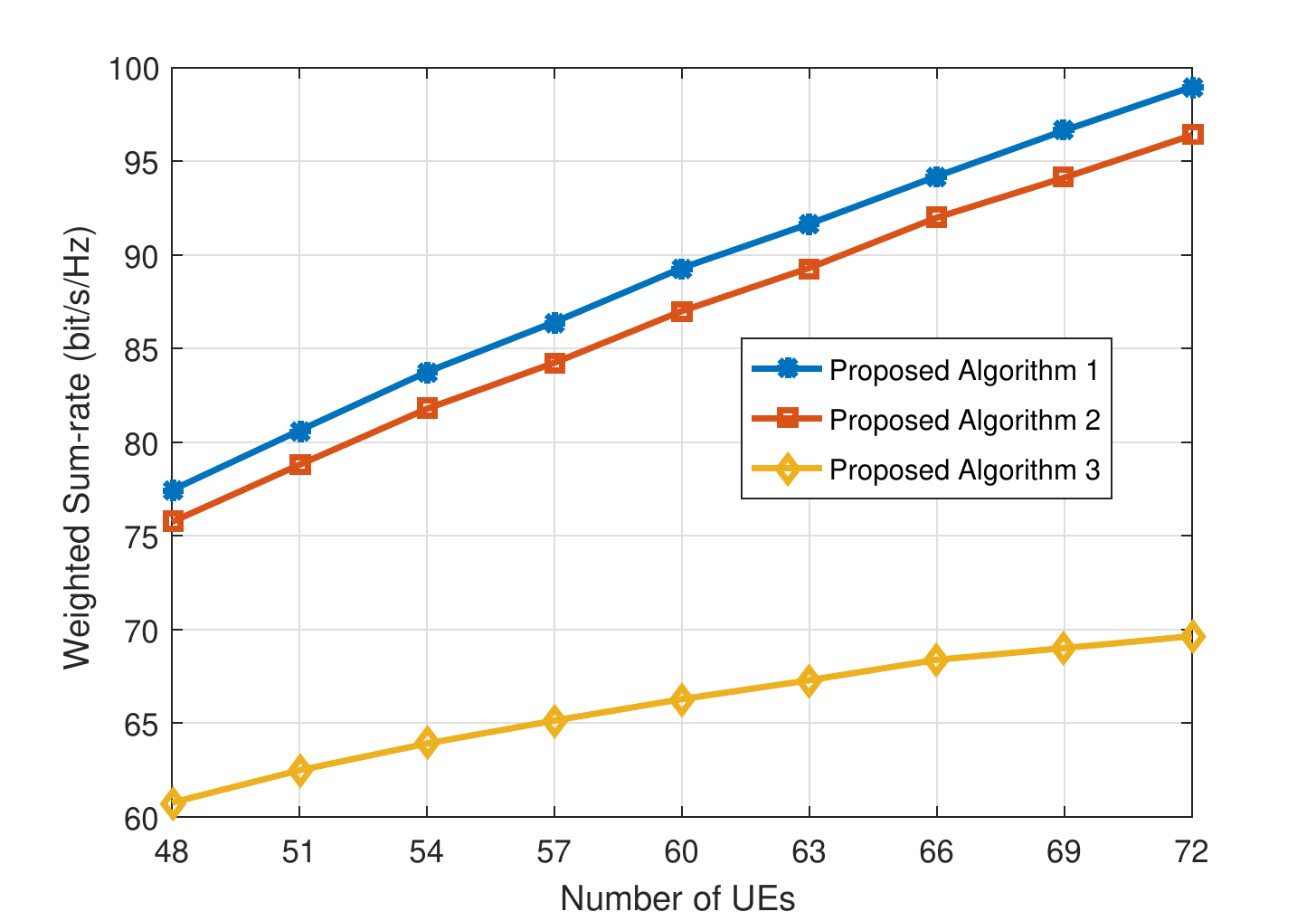}
\caption {The influence of the number of UEs on the sum rate.}
\label{sim4}
\end{figure}

Fig. \ref{sim4} shows the influence of the number of UEs $K$ on the sum rate when the number of BS antenna $N_t$ is fixed. It is found that the performance of the three proposed algorithms improves as $K$ increases. This is because more UEs can leverage the spatial multiplexing gain offered by the multiple-antenna BS. However, as $K$ increases, the co-channel interference also sharply increases. Under this condition, the first and second algorithms have more degrees of freedom to mitigate the interference, and thus their performance improves continuously. For the third one, there exists high residual interference after beamspace beamforming at the BS and SIC at the UEs. As a result, the sum rate of the third one increases slowly and becomes saturated fast.

\begin{figure}[h] \centering
\includegraphics [width=0.45\textwidth] {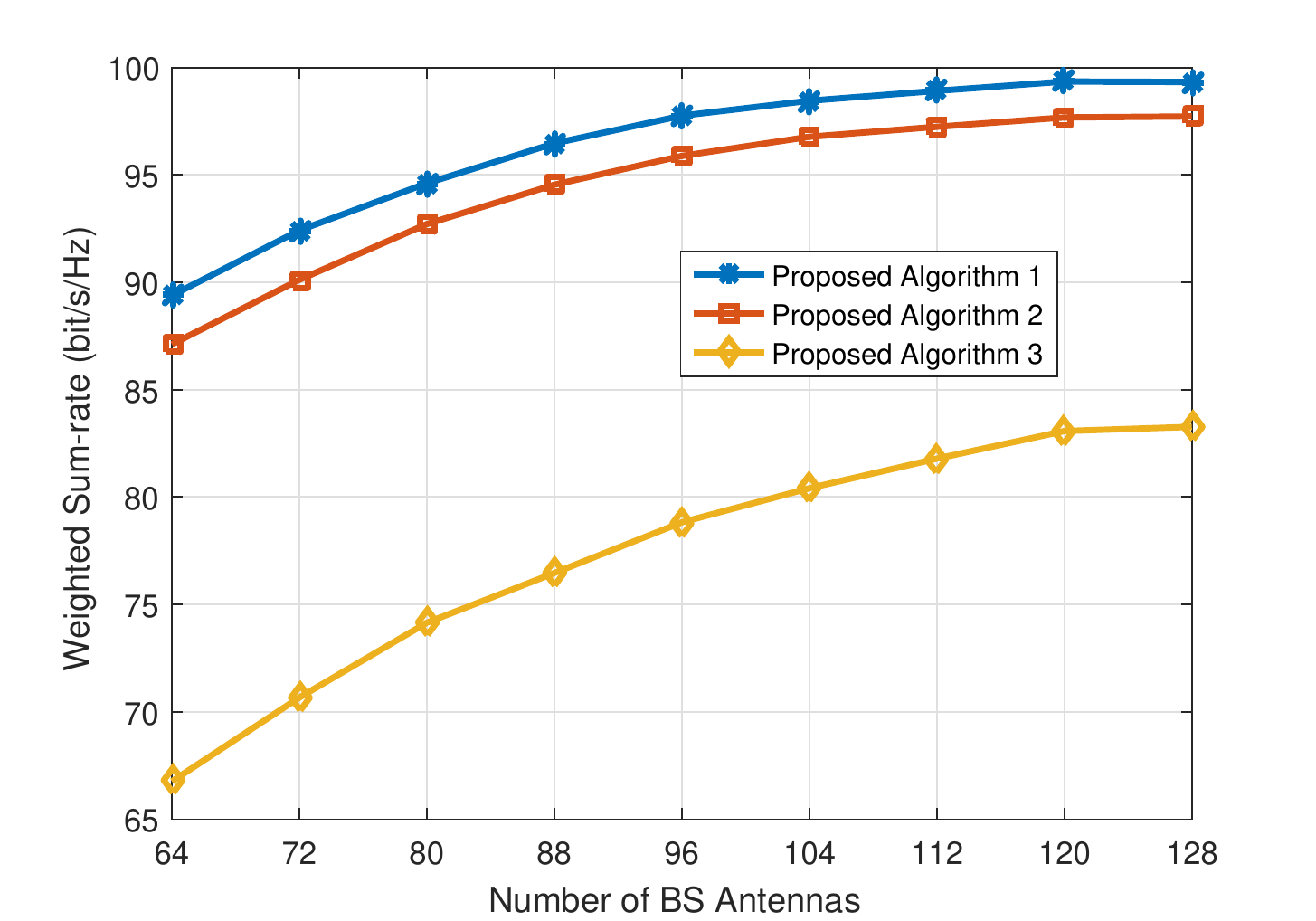}
\caption {The impact of the number of BS antennas on the sum rate.}
\label{sim5}
\end{figure}

Fig. \ref{sim5} investigates the impact of the number of BS antenna $N_t$ on the sum rate for a given number of UEs. As is intuitively observed, a large number of BS antennas can provide a high angular resolution in the beamspace, and hence designs highly accurate transmit beams for interference mitigation and signal enhancement. As a result, the sum rates of the three proposed algorithms improve as the number of BS antennas increases. Therefore, we can improve the performance of massive access by simply adding the BS antennas, which is a major advantage of the 5G BS with a large-scale antenna array. However, it is found that the performance gain by adding the BS antennas becomes small gradually. This is because there is nearly one UE in a cluster when the angular resolution is high enough.

\begin{figure}[h] \centering
\includegraphics [width=0.45\textwidth] {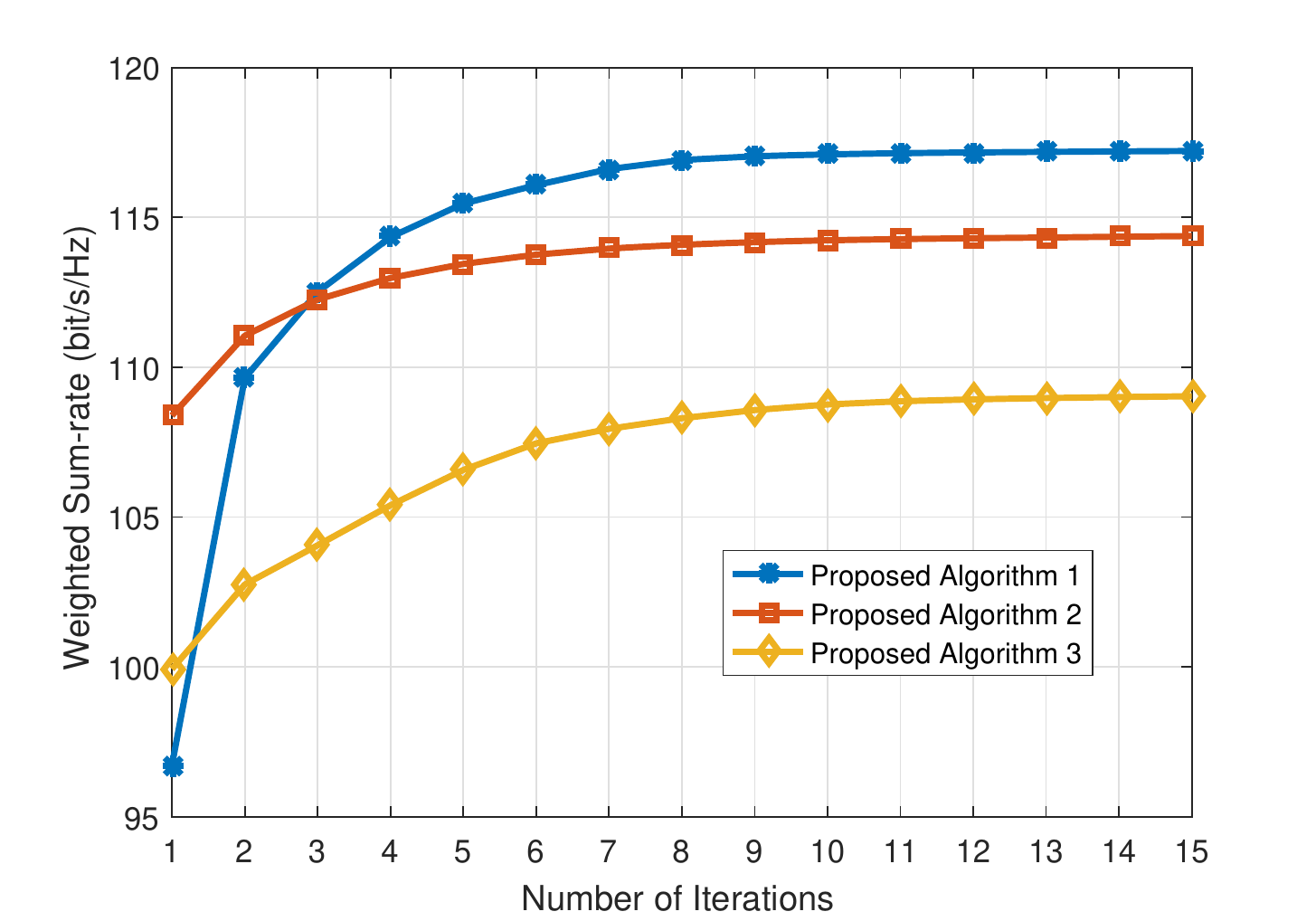}
\caption {The convergence behavior of the proposed three algorithms.}
\label{sim6}
\end{figure}

Finally, we check the convergence behavior of the three proposed algorithms in Fig. \ref{sim6}. It is seen that all the algorithms converge after no more than 10 times iterations. As discussed earlier, the beam design is a linear combination of multiple base beams with the power obtained by iterations. Thus, the proposed algorithms have low computational complexity and fast convergence behavior.

\section{Conclusion}
In this paper, we designed a massive access framework for the cellular IoT by making use of the characteristics of beamspace. Especially, we proposed to adopt non-orthogonal transmit beams to improve the performance of massive access systems. Three non-orthogonal beam design algorithms with different system performance and computation complexity were presented with the purpose of providing feasible solutions for the cellular IoT with distinct requirements. Simulation results have confirmed that the proposed non-orthogonal beamspace multiple access schemes can significantly improve the performance compared to the baseline ones.

\begin{appendices}
\section{The Proof of Theorem 1}
Prior to proving Theorem 1, we first provide the following lemma \cite{BDMA1}:

\emph{Lemma 2}: If $\mathbf{A}$, $\mathbf{B}$, and $\mathbf{X}$ are symmetric positive semi-definite matrices and $\mathbf{A}-\mathbf{B}\succeq\mathbf{0}$, the matrix function $f(\mathbf{X})$ in below is concave with respect to $\mathbf{X}$.
\begin{equation}
f(\mathbf{X})=\log_2\det(\mathbf{I}+\mathbf{A}\mathbf{X})-\log_2\det(\mathbf{I}+\mathbf{B}\mathbf{X}).\label{app1}
\end{equation}

\begin{figure*}
\begin{eqnarray}
r_{m,n}&=&\log_2\left(\frac{\sum\limits_{j=1,j\neq m}^{M}\sum\limits_{i=1}^{N_j}|\bar{\mathbf{h}}_{m,n}^H\mathbf{v}_{m,n,j,i}|^2+\sum\limits_{i=1}^{n}|\bar{\mathbf{h}}_{m,n}^H\mathbf{v}_{m,n,m,i}|^2+1}
{\sum\limits_{j=1,j\neq m}^{M}\sum\limits_{i=1}^{N_j}|\bar{\mathbf{h}}_{m,n}^H\mathbf{v}_{m,n,j,i}|^2+\sum\limits_{i=1}^{n-1}|\bar{\mathbf{h}}_{m,n}^H\mathbf{v}_{m,n,m,i}|^2+1}\right)\nonumber\\
&=&\log_2\det\left(\bar{\mathbf{h}}_{m,n}\bar{\mathbf{h}}_{m,n}^H\left(\sum\limits_{j=1,j\neq m}^{M}\sum\limits_{i=1}^{N_m}\mathbf{V}_{m,n,j,i}+\sum\limits_{i=1}^{n}\mathbf{V}_{m,n,m,i}\right)+\mathbf{I}\right)\nonumber\\
&&-\log_2\det\left(\bar{\mathbf{h}}_{m,n}\bar{\mathbf{h}}_{m,n}^H\left(\sum\limits_{j=1,j\neq m}^{M}\sum\limits_{i=1}^{N_m}\mathbf{V}_{m,n,j,i}+\sum\limits_{i=1}^{n-1}\mathbf{V}_{m,n,m,i}\right)+\mathbf{I}\right),\label{app2}
\end{eqnarray}
\begin{eqnarray}
\mathbb{E}\{r_{m,n}\}&\leq&\log_2\det\left(\mathbb{E}[\bar{\mathbf{h}}_{m,n}\bar{\mathbf{h}}_{m,n}^H]\left(\sum\limits_{j=1,j\neq m}^{M}\sum\limits_{i=1}^{N_m}\mathbf{V}_{m,n,j,i}+\sum\limits_{i=1}^{n}\mathbf{V}_{m,n,m,i}\right)+\mathbf{I}\right)\nonumber\\
&&-\log_2\det\left(\mathbb{E}[\bar{\mathbf{h}}_{m,n}\bar{\mathbf{h}}_{m,n}^H]\left(\sum\limits_{j=1,j\neq m}^{M}\sum\limits_{i=1}^{N_m}\mathbf{V}_{m,n,j,i}+\sum\limits_{i=1}^{n-1}\mathbf{V}_{m,n,m,i}\right)+\mathbf{I}\right)\nonumber\\
&=&\log_2\det\left(\sum\limits_{j=1,j\neq m}^{M}\sum\limits_{i=1}^{N_m}\mathbf{V}_{m,n,j,i}+\sum\limits_{i=1}^{n}\mathbf{V}_{m,n,m,i}+\mathbf{I}\right)\nonumber\\
&&-\log_2\det\left(\sum\limits_{j=1,j\neq m}^{M}\sum\limits_{i=1}^{N_m}\mathbf{V}_{m,n,j,i}+\sum\limits_{i=1}^{n-1}\mathbf{V}_{m,n,m,i}+\mathbf{I}\right),\label{app3}
\end{eqnarray}
\begin{eqnarray}
R_{\mathrm{ub}}&=&\sum\limits_{m=1}^{M}\sum\limits_{n=1}^{N_m}\alpha_{m,n}\left(\log_2\det\left(\sum\limits_{j=1,j\neq m}^{M}\sum\limits_{i=1}^{N_m}\mathbf{V}_{m,n,j,i}+\sum\limits_{i=1}^{n}\mathbf{V}_{m,n,m,i}+\mathbf{I}\right)\right.\nonumber\\
&&\left.-\log_2\det\left(\sum\limits_{j=1,j\neq m}^{M}\sum\limits_{i=1}^{N_m}\mathbf{V}_{m,n,j,i}+\sum\limits_{i=1}^{n-1}\mathbf{V}_{m,n,m,i}+\mathbf{I}\right)\right)\nonumber\\
&=&\sum\limits_{m=1}^{M}\sum\limits_{n=1}^{N_m}\sum\limits_{c=1}^{N_t}\alpha_{m,n}\left(\log_2\left(\left(\sum\limits_{j=1,j\neq m}^{M}\sum\limits_{i=1}^{N_j}s_{j,i,c}p_{j,i,c}+\sum\limits_{i=1}^{n}s_{m,i,c}p_{m,i,c}\right)\eta_{m,n,c}+1\right)\right.\nonumber\\
&&\left.-\log_2\left(\left(\sum\limits_{j=1,j\neq m}^{M}\sum\limits_{i=1}^{N_j}s_{j,i,c}p_{j,i,c}+\sum\limits_{i=1}^{n-1}s_{m,i,c}p_{m,i,c}\right)\eta_{m,n,c}+1\right)\right).\label{app4}
\end{eqnarray}
\end{figure*}

According to the definition, the achievable rate of the UE$_{m,n}$ in (\ref{eqn13}) can be transformed as (\ref{app2}) at the top of the next page, where $\mathbf{v}_{m,n,j,i}=\mathbf{\Lambda}_{m,n}^\frac{1}{2}\mathbf{P}_{j,i}^\frac{1}{2}\mathbf{s}_{j,i}$ and $\mathbf{V}_{m,n,j,i}=\mathbf{v}_{m,n,j,i}\mathbf{v}_{m,n,j,i}^H$. Eq. (\ref{app2}) holds true due to the fact that $\det(\mathbf{I}+\mathbf{AB})=\det(\mathbf{I}+\mathbf{BA})$. According to Lemma 2, $r_{m,n}$ in (\ref{app2}) is a concave function of $\|\bar{\mathbf{h}}_{m,n}\|^2$. Thus, applying the Jensen's inequality yields Eq. (\ref{app3}).
Then, the upper bound of the weighted sum of the ergodic rates can be written as Eq. (\ref{app4}), where (\ref{app4}) follows the fact that $\mathbf{V}_{m,n,j,i}$ is a diagonal matrix. The proof completes.
\end{appendices}

%\end{spacing}
\end{document}